\documentclass[useAMS,usenatbib]{mn2e}
\usepackage{graphicx}
\usepackage{rotating}
\usepackage{lscape}
\usepackage{makeidx}
\usepackage{listings}
\usepackage{longtable}
\usepackage{array}
\usepackage{aas_macros}
\usepackage{subfigure}
\usepackage{times}
\usepackage{hyperref}

\title[H-ATLAS: radio-loud/quiet quasars]{\emph{Herschel}-ATLAS\thanks{{\it Herschel} is an ESA
    space observatory with science instruments provided by
    European-led Principal Investigator consortia and with important
    participation from NASA}: Far-infrared properties of radio-loud and radio-quiet quasars}
\author[Kalfountzou et al.]
{E. Kalfountzou$^{1}$\thanks{Email: e.kalfountzou@herts.ac.uk}, J.A. Stevens$^{1}$,  M.J. Jarvis$^{2,3}$,  M.J. Hardcastle$^{1}$,  D.J.B. Smith$^{1}$, 
\newauthor N. Bourne$^{4}$, L. Dunne$^{5}$, E. Ibar$^{6}$, S. Eales$^{7}$, R. J. Ivison$^{8,12}$, S. Maddox$^{9}$, M.W.L. Smith$^{10}$, 
\newauthor E. Valiante$^{10}$, G. de Zotti$^{11}$
 \\
\footnotesize
$^{1}$Centre for Astrophysics, Science \& Technology Research Institute, University of Hertfordshire, Hatfield, Herts, AL10 9AB, UK\\
$^{2}$Oxford Astrophysics, Denys Wilkinson Building, University of Oxford, Keble Rd, Oxford OX1 3RH\\
$^{3}$Physics Department, University of the Western Cape, Cape Town, 7535, South Africa\\
$^{4}$School of Physics \& Astronomy, University of Nottingham, University Park, Nottingham NG7 2RD, UK \\
$^{5}$Dept Physics and Astronomy, University of Canterbury, Private Bag 4800, Christchurch, New Zealand \\
$^{6}$Instituto de F\'isica y Astronom\'ia, Universidad de Valpara\'iso, Avda. Gran Breta\~na 1111, Valpara\'iso, Chile\\
$^{7}$School of Physics and Astronomy, Cardiff University, Queen's Buildings, 5 The Parade, CF24 3AA, Cardiff, UK \\
$^{8}$Institute for Astronomy, University of Edinburgh, Royal Observatory, Blackford Hill, Edinburgh EH9 3HJ \\
$^{9}$Department of Physics and Astronomy, University of Canterbury, Private Bag 4800, Christchurch, New Zealand \\
$^{10}$School of Physics and Astronomy, Cardiff University, Queen's Buildings, The Parade, Cardiff CF24 3AA, UK \\
$^{11}$INAF-Osservatorio Astronomico di Padova, Vicolo Osservatorio 5, I-35122 Padova, Italy, and SISSA, Via Bonomea 265, I-34136 Trieste, Italy \\
$^{12}$European Southern Observatory, Karl Schwarzschild Strasse 2, D-85748 Garching, Germany}

\begin{document}
\date{Received Month dd, yyyy; accepted Month dd, yyyy}
\pagerange{\pageref{firstpage}--\pageref{lastpage}} \pubyear{2012}
\maketitle
\label{firstpage}

\begin{abstract}
\\
We have constructed a sample of radio-loud and radio-quiet quasars from the Faint Images Radio Sky at Twenty-one centimetres (FIRST) and the Sloan Digital Sky Survey Data Release 7 (SDSS DR7), over the H-ATLAS Phase 1 Area ($9^{h}$, $12^{h}$ and $14.5^{h}$). Using a stacking analysis we find a significant correlation between the far-infrared luminosity and 1.4-GHz luminosity for radio-loud quasars. Partial correlation analysis confirms the intrinsic correlation after removing the redshift contribution while for radio-quiet quasars no partial correlation is found. Using a single-temperature grey-body model we find a general trend of lower dust temperatures in the case of radio-loud quasars comparing to radio-quiet quasars. Also, radio-loud quasars are found to have almost constant mean values of dust mass along redshift and optical luminosity bins. In addition, we find that radio-loud quasars at lower optical luminosities tend to have on average higher FIR and 250-$\mu$m luminosity with respect to radio-quiet quasars with the same optical luminosites. Even if we use a two-temperature grey-body model to describe the FIR data, the FIR luminosity excess remains at lower optical luminosities. 
These results suggest that powerful radio jets are associated with star formation especially at lower accretion rates.


\end{abstract}


\section{INTRODUCTION}

\subsection{AGN and star-formation connection}

Star formation and Active Galactic Nucleus (AGN) activity play important roles in the formation and evolution of galaxies. Over the past two decades a significant number amount of evidence has demonstrated the close connection between AGNs and their hosts. A tight correlation exists between black hole and galaxy bulge masses \citep[e.g.][]{Boyle1998,Ferrarese2000,McLure2001,Merloni2004}. In addition, the evolutionary behaviour of AGN shows a strong correlation with luminosity: the space density of luminous AGN peaks at $z\sim2$, while for lower luminosity AGN it peaks at $z\sim1$ \citep[e.g.][]{Hasinger2005,Babic2007,Bongiorno2007,Rigby2011}. This so-called anti-hierarchical evolution is similar to the downsizing behaviour of galaxy star-formation activity \citep[e.g.][]{Cowie1996,Fontanot2009} which, in some cases, is associated with the decline in frequency of major mergers \citep[e.g.][]{Treister2012}. Although AGN activity and star formation in galaxies do appear to have a common triggering mechanism, recent studies do not find strong evidence that the presence of AGN affects the star-formation process in the host galaxy \citep[e.g.][]{Bongiorno2012,Feltre2013}.

Theoretical models suggest that these possible correlations arise through feedback processes between the galaxy and its accreting black hole. Such regulation has been shown to be important in large cosmological simulations \citep[e.g.][]{DiMatteo2005,Springel2005, Croton2006}. In general these can take two forms, AGN-winds (often referrred to as quasar-mode) which comprise wide-angle, sub-relativistic outflows and tend to be driven by the radiative output of the AGN, and jets (often referred to as radio-mode), which are relativistic outflows with narrow opening angles that are launched directly from the accretion flow itself.  In the case of quasar-mode the objects are accreting rapidly, at near their Eddington rate and their radiation can couple to the gas and dust in the interstellar medium, driving winds that may shut down further accretion onto the black hole or even drive material out of the galaxy, thereby quenching star formation \citep[e.g.][]{DiMatteo2005}. Although there is no compelling evidence for AGN feedback quenching star formation, there is mounting evidence for quasar-driven outflows \citep[e.g.][]{Maiolino2012}.However recent surveys find little evidence that X-ray luminous AGN quench star formation (\citealp{Harrison2012} cf. \citealp{Page2012}). Similarly, the radio-mode  and the role of radio-loud AGN and their jets in the evolution of galaxies has been studied intensively suggesting that jets can have positive as well as negative feedback on star-formation rates with the observational consensus being mixed. Certainly, some studies advocate that radio-jets effectively suppress or even quench star formation \citep[e.g.][]{Best2005,Croton2006,Best2012,Karouzos2013,Chen2013} by warming-up and ionizing the interstellar medium (ISM) which leads to less efficient star formation, or through direct expulsion of the molecular gas from the galaxy, effectively removing the ingredient for stars to form \citep[e.g.][]{Nesvadba2006,Nesvadba2011}. On the other hand, positive feedback can enhance star formation which could be explained by shocks driven by the radio-jets in the ISM that compress it and eventually lead to enhanced star-formation efficiency \citep[e.g.][]{SilkNusser10,Kalfountzou2012,Gaibler2012,Best2012}.

It is therefore apparent, that although some form of feedback is needed to explain the observational results supporting co-evolution of central spheroids and their galaxies, much still remains unclear. Radio-loud and radio-quiet quasars provide ideal candidates for the study of star formation in powerful AGN under the presence of jets or otherwise. Indeed, optically selected radio-loud quasars are found to have enhanced star formation at lower luminosities using optical spectral feature as a diagnostic \citep{Kalfountzou2012}. The latter result raises the question of why such an effect is not seen at high radio power and/or AGN activity which could be explained under the assumption of a dominant mechanical feedback at low Eddington luminosities, in which case this would plausibly be the major source of positive feedback.

However, spectral diagnostics are not immune to AGN contamination and optical diagnostics, in particular, are susceptible to the effects of reddening. Indeed, the measurement of the star-formation activity in the host galaxy is difficult, mainly due to contamination by the AGN. Many studies have attempted to determine the star-formation activity in quasar host galaxies using optical colours \citep[e.g.][]{Sanchez2004} or spectroscopy \citep[e.g.][]{Trichas2010,Kalfountzou2011,Trichas2012}. or X-ray selection \citep[e.g.][]{Comastri2003,Treister2011}. In addition, AGN emission can outshine both the ultra-violet (UV) and optical emission from young stars. By contrast, the far-infrared (FIR) emission is shown to be dominated by emission from dust in the host galaxy, except in the most extreme cases \citep[e.g.][]{Netzer2007,Mullaney2011}, and to be a proxy of its star formation activity that is largely uncontaminated by the AGN \citep[e.g.][]{Haas2003,Hatziminaoglou2010}.

\subsection{Radio-loud and radio-quiet quasars}

A property of quasars is the existence of radio-loud and radio-quiet populations. One of the more controversial topics in studies of these objects is whether these radio-loud and radio-quiet quasars form two physically distinct populations of objects. Radio-loud quasars are often defined to be the subset of quasars with a radio-loudness satisfying $R_{i}>10$, where $R_{i}=L({\rm 5GHz})/L({\rm 4000\AA})$ \citep{Kellermann1989} is the ratio of monochromatic luminosities measured at (rest frame) 5~GHz and 4000~\AA. Radio-quiet quasars must minimally satisfy $R_{i}\leq10$. However, even radio-quiet quasars quasars can be detected as radio sources \citep{Kellermann1989}. This has led to two opposing views of the radio-loudness distribution which have long been debated. The first is that the radio-loudness distribution is bimodal \citep[e.g.][]{Kellermann1989,Miller1990,Ivezic2002}. The other is that the distribution is continuous with no clear dividing line \citep[e.g.][]{Cirasuolo2003,LaFranca2010,Singal2011,Singal2013,Bonchi2013}. Typically, optically selected radio-loud quasars are only a small fraction, $\sim$10-20 per cent, of all quasars (e.g. \citealp{Ivezic2002} but see also \citealp{Richards2006} with a small radio-loud fraction of 3 per cent), with this fraction possibly varying with both optical luminosity and redshift \citep{Jiang2007}. In contrast, X-ray selected samples show lower fractions of radio-loud AGN $<5$ per cent \citep[e.g.][]{Donley2007,LaFranca2010}. However, many low-power radio sources in these samples might be star formation-driven \citep[e.g.][]{Massardi2010}. X-ray selections overall probe much higher (or complete) portions of the AGN populations than optical ones. This may affect the comparison of same subsamples (i.e., radio-loud) selected with different methods. Radio-loud quasars usually reside in very massive galaxies and have typically a lower optical or X-ray output at given stellar mass (i.e. lower $L/L_{\rm Edd}$ at given $L_{\rm Edd}$, \citealp{Sikora2007}) compared to radio-quiet quasars. This means that an $L_{X}$-limited sample will have a lower radio-loud quasars fraction, compared to a mass-limited sample. However, in the case of a strictly limited selection of X-ray-Type I AGN, then possibly the subsamples of radio-loud AGN might end up being more comparable to optical ones.

While a definitive physical explanation of this dichotomy remains elusive, a large number of models have been put forward to explain it. Both types of quasars are likely powered by similar physical mechanisms \citep[e.g.][]{Urry1995,Shankar2010}, but their radio loudness has been shown to be anti-correlated with accretion rate onto their central supermassive black holes \citep[e.g.][]{Fernandes2011}. Additionally, it has been demonstrated that, relative to radio-quiet quasars, radio-loud quasars are likely to reside in more massive host galaxies (\citealp{Kukula2001,Sikora2007}). However, \cite{Dunlop2003} found that spheroidal hosts become more prevalent with increasing nuclear luminosity such that, for nuclear luminosities ${\rm M_{V} < -23.5}$, the hosts of both radio-loud and radio-quiet AGN are virtually all massive ellipticals.

Along with the idea of different host galaxies it has been found that radio-loud quasars require more massive central black holes than radio-quiet quasars (e.g. \citealt{Dunlop2003, McLure2004}; see also \citealp{Shankar2010}, who finds this to be redshift dependent) and it has also been suggested that radio-loud quasars host more rapidly spinning black holes than radio-quiet quasars (e.g. \citealt{BlandfordZnajek1977, Punsly1990, Wilson1995, Sikora2007, Fernandes2011}; but see also \citealp{Garofalo2010}). The low radio-loud fraction also suggests a change in jet occurrence rates among active super-massive black holes at low luminosities. This could be linked to changes in the Eddington fraction, evolutionary state of the black hole, or the host galaxy mass, evolutionary state, or environment.
Recently, \cite{Falder2010} showed that radio-loud AGN appear to be found in denser environments than their radio-quiet counterparts at $z\sim 1$, in contrast with previous studies at lower redshifts (e.g. \citealp{McLure2001}). However the differences are not large and may be partly explained by an enhancement in the radio emission due to the confinement of the radio jet in a dense environment (e.g. \citealp{Barthel1996}).

If the radio-loudness is due to the physics of the central engine and how it is fueled, and the environment plays a relatively minor role, the quasar properties may be connected with the star formation in their host galaxies \citep[e.g.][]{Herbert2010,Hardcastle2013}. On the one hand, AGN feedback could be stronger in the case of the radio-loud quasars due to their higher black hole masses and therefore potentially stronger radiation field, reducing the star-formation rate compared to radio-quiet quasars; on the other hand radio jets could increase the star-formation activity by compressing the intergalactic medium \citep[e.g.][]{Croft2006, SilkNusser10}.

\subsection{This work}

With the \textit{Herschel} Space Observatory \citep{Pilbratt2010} we are able to measure the FIR emission of AGN host galaxies and hence the cool-dust emission. \textit{Herschel} offers an ideal way of measuring the instantaneous star-formation rate (SFR) of AGN \citep[e.g.][]{Bonfield2011}. Until \textit{Herschel}, hot dust emission has typically been determined from Spitzer data at near/mid-infrared wavelengths, but emission from the torus can also contribute at these bands, especially in the case of quasars. With \textit{Herschel} we are able to determine the level of cool dust emission in AGN, providing a detailed picture of how the full SEDs of AGN change as a function of luminosity, radio-loudness and redshift. Under these circumstances, \textit{Herschel} provides a good tool to study the star formation and AGN activity in a special type of AGN: quasars. We are also able to study the star formation in different types of quasars (e.g. radio-loud and radio-quiet quasars) and thus to say how it might be affected by the presence of powerful radio jets.

The paper is structured as follows. In section \ref{data} we discuss the selection of the sample and the observations we have used. 
In section \ref{sec:FIR} we describe the statistical methods and the models we have used in order to estimate the FIR parameters (e.g. FIR luminosity, dust temperature, dust mass) of our sample. Here we also present the results of the comparison of the FIR parameters between the radio-loud and radio-quiet quasars. Finally, in sections \ref{sec:discussion} and \ref{sec:conclusions}, we explore the general conclusions that can be drawn from our results. 

Throughout the paper we use a cosmology with $H_{0}=70{~\rm km~s^{-1}~Mpc^{-1}}$, $\Omega_{m}=0.3$ and $\Omega_{\Lambda}=0.7$.

\section{SAMPLE DEFINITION AND MEASUREMENTS} \label{data}

\subsection{The data}

In this section we describe the data used throughout this paper.

\begin{figure*}
\includegraphics[scale=0.43]{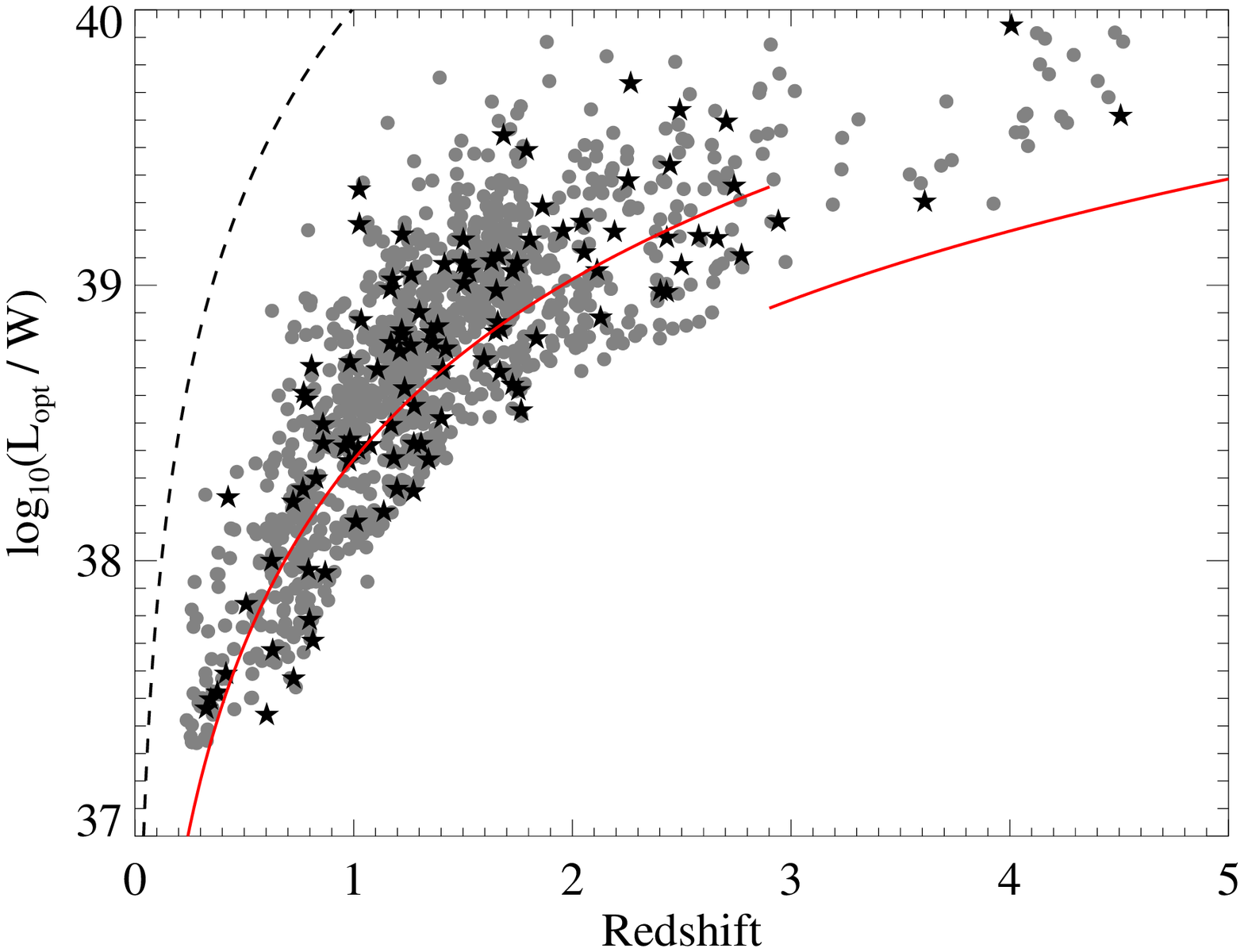}
\includegraphics[scale=0.43]{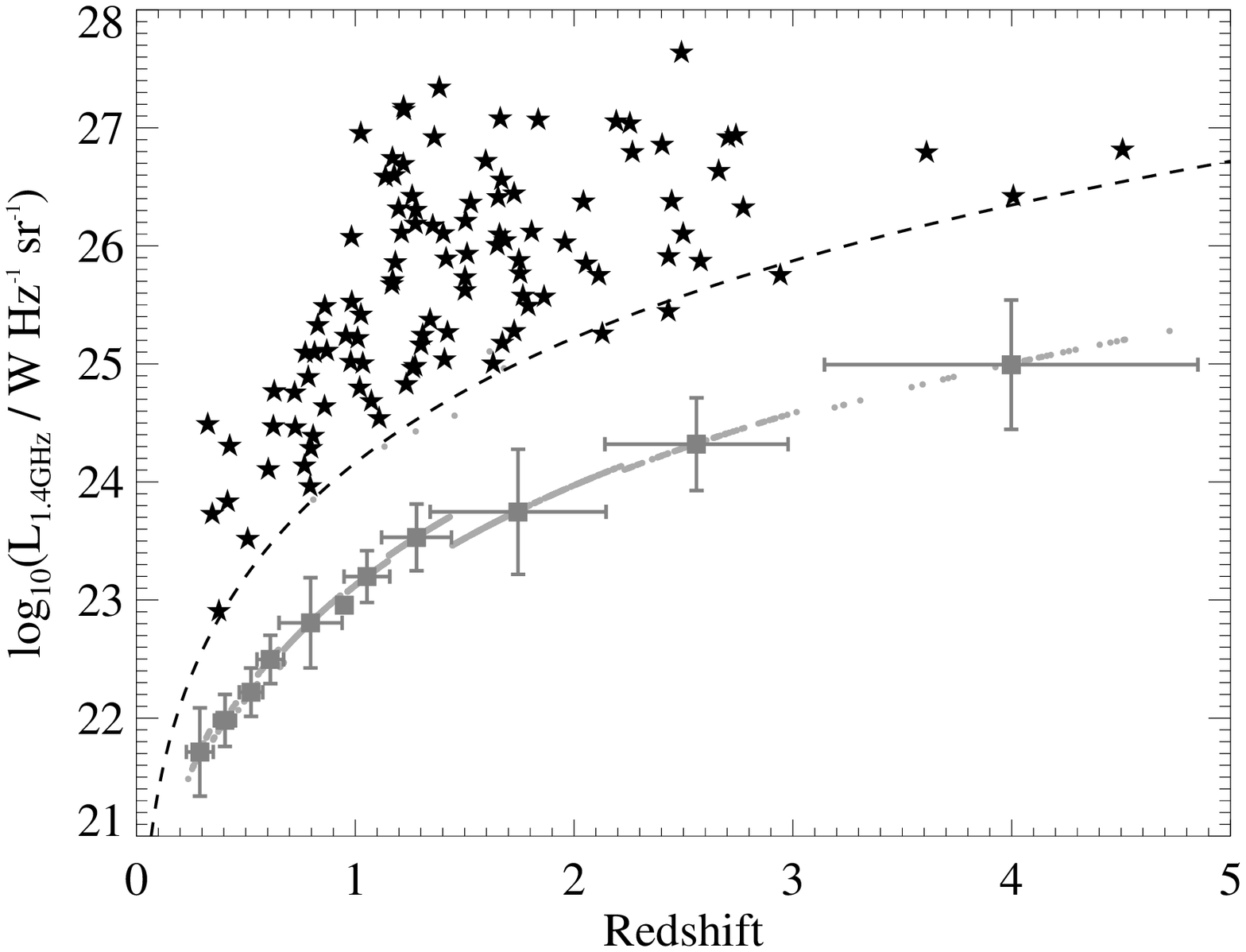}
\caption{Left: Optical luminosity of the radio-loud (black stars) and radio-quiet (grey circles) samples as a function of redshift. The red lines show the corresponding ${\rm M}_{i}$ for $i=19.1$ $(z<2.9)$ and $i=20.2$ ($z>2.9$), respectively, and the black dashed line shows the equivalent for $i=15$ (the bright limit for SDSS quasar targets; \citealp{2011ApJS..194...45S}). Right: Radio luminosity as a function of redshift. The mean values and the errors for undetected quasars are represented by large grey circles. The dashed line corresponds to the nominal $5\sigma$ flux cut-off of FIRST, i.e. $1.0~{\rm mJy}$.}
\label{fig:LvsZ}
\end{figure*}

\begin{enumerate}
\item Radio source catalogues and images from the Faint Images of the Radio Sky at Twenty-one centimetres (FIRST; \citealp{Becker1995}) survey and NRAO NLA Sky Survey (NVSS; \citealp{1998AJ....115.1693C}). Both cover the entire H-ATLAS \citep{Eales2010} Phase 1 area. To check the possibility of non-thermal contamination in the \textit{Herschel} bands, we also cross matched our sample with the Giant Metrewave Radio Telescope (GMRT) catalogue of \cite{Mauch2013}, who have imaged the majority of the Phase 1 area at 325 MHz, in order to estimate the radio spectral index for the radio-loud sample.

\item Point spread function (PSF) convolved, background subtracted images of the H-ATLAS Phase 1 fields at wavelengths of 100, 160, 250, 350 and 500 $\mu$m, provided by the Photodetector Array Camera \& Spectrometer (PACS; \citealp{Poglitsch2010}) and the Spectral and Photometric Imaging REceiver (SPIRE; \citealp{2010A&A...518L...3G}) instruments on the \textit{Herschel Space Observatory}. The Phase 1 area consists of three equatorial strips centred at $9^{h}$, $12^{h}$ and $14.5^{h}$. Each field is approximately $12^{\rm o}$ in RA by $3^{\rm o}$ in Dec ($6^{\rm o}$ by $3^{\rm o}$ for the $12^{h}$ field). The construction of these maps is described in detail by \cite{2011MNRAS.415..911P} for SPIRE. From these maps, a catalogue of the FIR sources was generated \citep{2011MNRAS.415.2336R}\footnote{The cited paper is for the SV data release, but the same processing techniques were used to create the catalogue for the Phase 1 area.}, which includes any source detected at 5$\sigma$ or better at any SPIRE wavelength. PACS fluxes were derived using apertures placed on the maps \citep{Ibar2010} at the locations of the 250 $\mu$m positions. The $5\sigma$ point source flux limits are 132, 121, 30.4, 36.9 and 40.8 mJy, with beam sizes ranging from 9 to 35 arcsec FWHM in the 100, 160, 250, 350 and 500 $\mu$m bands, respectively.

\item Redshift and optical magnitudes from the Sloan Digital Sky Survey Data Release 7 (SDSS DR7) Quasar Catalogue \citep{2010AJ....139.2360S} which provides the most reliable classification and redshift of SDSS quasars with absolute $i'-$band magnitudes brighter than -22.
\end{enumerate}

We constructed a sample of radio-detected quasars in the FIRST field with optical magnitudes and redshifts from SDSS DR7. A matching radius $r\leq5$ arcsec is used to identify the compact radio sources while a larger radius of $30$ arcsec is used for extended sources. With this method we found 144 quasars with matching radius less than $5''$ and 3 extended quasars.

In order to check that the radio maps from the FIRST survey do not miss a significant fraction of extended emission around the quasars, we also cross-correlate the optical positions with NVSS. For the undetected quasars in FIRST we used a stacking analysis to estimate their flux densities following \cite{2007ApJ...654...99W}, where they quantified the systematic effects associated with stacking FIRST images and examined the radio properties of quasars from the SDSS by median-stacking radio maps centered on optical position of these quasars. More details of the cross-matching, the stacking analysis and the radio-loudness parameter are described by \cite{Kalfountzou2012}. 

A total of 1,618 quasars (141 radio-loud and 1,477 radio-quiet quasars) are found in the H-ATLAS Phase 1 field based on their optical positions. For this sample, we have investigated how many quasars are significantly detected in the H-ATLAS catalogue at the 5$\sigma$ level. Cross-matching with the H-ATLAS Phase 1 Catalogue applying a likelihood ratio technique \citep{2011MNRAS.416..857S} yielded 146 ($\sim9$ per cent) counterparts with a reliability $R>0.8$. Among the 146 counterparts 9 are radio-loud quasars ($\sim7$ per cent of the radio-loud population). A similar percentage was found by \cite{Bonfield2011}. Comparing the detected samples of radio-loud and radio-quiet quasars by applying a K-S test gives a null hypothesis of $p=0.07$, $p=0.11$, $p=0.08$, $p=0.11$ and $p=0.14$ for 100, 160 $\mu$m PACS and 250, 350 and 500 $\mu$m SPIRE bands.

Since the radio-loud sample includes sources with high radio flux density we also investigated the possibility of synchrotron contamination, which is not associated with star formation, to the FIR flux densities. The method we are using to estimate the synchrotron contamination is described in Appendix A. We have found that out of the 141 objects in our radio-loud quasar sample, 21 radio-loud quasars have significant non-thermal contamination in their FIR emission. These objects have been removed from our sample. We have also found that 27 sources are possible candidates for strong contamination using an upper limit for their radio spectral index. These sources have been also removed from the sample.

We then compare the distribution in $z$ and $L_{\rm opt}$ of radio-loud and radio-quiet quasars and force the two subsamples to have the same $L_{\rm opt}$ and $z$ distribution by randomly removing radio-quiet quasars from our parent sample. Running a K-S test on these samples we find the distribution of the two populations in the optical luminosity - redshift plane is similar. A Kolmogorov-Smirnov test (K-S test) applied to the optical luminosity gives a result that corresponds to a probability, $p=0.69$ under the null hypothesis (i.e. they are statistically indistinguishable) while the K-S test to the redshift gives $p=0.75$. A 2-d K-S test on the redshift and optical luminosity for both samples returns $p=0.58$. We can therefore assume the populations are matched in optical luminosity and $z$. This process provides a radio-optical catalogue of quasars with spectroscopic redshift up to $z\sim 5$. Fig.~\ref{fig:LvsZ} shows the optical luminosity - redshift and the radio luminosity - redshift plots for the final sample of 93 radio-loud and 1,007 radio-quiet quasars. We have randomly removed 470 radio-quiet quasars from our original sample in order to match the two populations into $z$ and $L_{\rm opt}$.

The optical luminosity was measured using the $i′$-band magnitude since redder passbands measure flux from the part of the spectrum relatively insensitive to recent star formation and also suffer less dust extinction. Since the $i$-band luminosity itself is expected to correlate with the AGN luminosity and is less sensitive to recent star-formation activity we use the optical luminosity as an AGN tracer. The rest-frame 1.4-GHz radio luminosities of the quasars were calculated from the FIRST 1.4-GHz flux density and the spectroscopic redshift, assuming a power law of $S_{\nu}\propto \nu^{-\alpha}$. The spectral index was measured using the FIRST and GMRT data. For the sources undetected by GMRT either a spectral slope of $\alpha = 0.71$ was used or the estimated spectral index using the nominal 5 mJy limit of the GMRT data (see Appendix).

\subsection{\emph{Herschel} flux measurements and stacked fluxes} \label{Herschel fluxes}

Due to the limited sample of SPIRE-detected quasars, especially the radio-loud quasars, we directly measure the FIR flux densities from the PSF-convolved images for all three H-ATLAS fields rather than just use the $5\sigma$ catalogues. For each of the quasars found inside the H-ATLAS Phase 1 field we derive the FIR flux densities in the two PACS and the three SPIRE bands directly from the background-subtracted, PSF-convolved H-ATLAS images. We take the flux density to be the value in the image at the pixel closest to the optical position of our targets. The errors are estimated from the centroid of the corresponding noise map including the confusion noise. In addition, the current H-ATLAS catalogue recommends including calibration errors of 10 per cent of the estimated flux for the PACS bands and 7 per cent for the SPIRE bands. The flux densities are background subtracted using a mean background value for each band. The mean background is estimated from 100,000 randomly selected pixels within the three H-ATLAS blank fields. 

To establish whether sources in the bins were significantly detected, we compared the flux measurements with the background flux distribution from 100,000 randomly selected position in the fields, following \cite{Hardcastle2010}. Using a K-S test, we can examine whether the flux densities are statistically distinguishable from those taken from randomly chosen positions, as a K-S test is not influenced by the non-Gaussian nature of the noise as a result of confusion. We found a distinguishable difference in all bands with K-S probability lower than $10^{-5}$. The mean background flux densities are $0.06\pm0.01$, $0.10\pm0.02$, $1.12\pm0.03$, $2.91\pm0.04$ and $0.51\pm0.03$~mJy at 100, 160, 250,350 and 500$\mu$m, respectively.

We have separated the samples in bins corresponding to redshift, radio luminosity and optical luminosity to investigate whether the far-infrared fluxes vary with those parameters. Within each bin we have estimated the weighted mean of the FIR background-subtracted flux densities in each \textit{Herschel} band. The mean values for each band are shown in Table~\ref{Table:mean_fluxes}. The errors have been determined by bootstrapping. The bootstrapped errors are determined by randomly selecting galaxies from within each bin and determining the median for this subsample. The K-S test results for the two populations and the Mann-Whitney (M-W) test results are also presented in Table~\ref{Table:KS_fluxes}. We find that there is no statistical difference between the FIR flux densities of radio-loud and radio-quiet quasars as a whole. However, separating the two populations into redshift and optical luminosity bins we find different results. With this division, we can see that at lower redshifts and/or lower optical luminosities the mean 350 $\mu$m and 500 $\mu$m flux densities for the radio-loud objects are significantly higher than for the radio-quiet ones at greater than the 3$\sigma$ level. 

\begin{table*}
\begin{center}
\caption{The radio-loud and radio-quiet quasars (RLQs and RQQs) FIR mean flux densities in the 100, 160, 250, 350 and 500 $\mu$m bandpasses. The two populations have been separated into redshift, radio luminosity and optical luminosity bins. The number of objects within each stack is also given.}
\centering
\begin{tabular} {c c c c c c c c }

\hline 

Class	&	$z$-range		&		$N$ per	bin	&	\multicolumn{5}{c}{Mean flux density (mJy)}		\\
		&	 	&		 	& 100 $\mu$m	&	160 $\mu$m	&		250 $\mu$m	&	350 $\mu$m	&	500 $\mu$m		\\

\hline

RLQs	&	$0.2-1.0$	&	24	&	$7.9\pm1.9$	&	$7.6\pm1.7$	&	$18.5\pm2.2$	&	$26.9\pm4.4$	&	$20.1\pm3.4$			\\
	&	$1.0-1.5$	&	30	&	$8.2\pm1.8$	&	$16.7\pm4.2$	&	$36.8\pm2.1$	&	$40.4.0\pm3.9$	&	$34.2\pm3.2$	\\
	&	$1.5-2.0$	&	21 	&	$4.1\pm1.4$	&	$7.1\pm1.3$	&	$17.3\pm2.3$	&	$23.0\pm2.2$	&	$18.5\pm2.7$		\\
	&	$2.0-5.0$	&	18	&	$2.6\pm1.4$	&	$5.6\pm1.9$	&	$18.3\pm2.3$	&	$23.5\pm2.1$	&	$21.3\pm2.8$ \\

	&			&		&			&			&			&			&	\\

RQQs	&	$0.2-1.0$	&	264		&	$7.3\pm0.5$		&	$9.9\pm0.7$	&	$21.2\pm1.0$	&	$20.0\pm0.7$	&	$12.3\pm0.6$ \\
	&	$1.0-1.5$	&	355		&	$5.1\pm0.6$		&	$8.2\pm1.0$		&	$20.3\pm1.6$	&	$21.2\pm1.0$	&	$13.7\pm0.6$ \\
	&	$1.5-2.0$	&	230		&	$3.9\pm0.4$		&	$9.5\pm0.5$	&	$18.7\pm0.8$	&	$21.2\pm0.7$	&	$14.7\pm0.7$	 \\
	&	$2.0-5.0$	&	158		&	$4.4\pm0.6$		&	$7.4\pm1.2$		&	$17.59\pm1.7$	&	$22.2\pm1.2$	&	$16.7\pm1.0$ \\
		
\hline
\hline

Class	&	$\log_{10}(L_{\rm 1.4}/{\rm W~Hz^{-1}})$	 &	$N$ per	bin	&	\multicolumn{5}{c}{Mean flux density (mJy)}		\\
		&					&	 &	100 $\mu$m	&	160 $\mu$m		&	250 $\mu$m	&	350 $\mu$m	&	500 $\mu$m		 \\

\hline

RLQs	&	$23.0-25.0$	&	20 &	$9.7\pm1.7$ &	$9.2\pm2.5$	&	$19.9\pm3.3$	&	$27.4\pm3.1$	&	$20.1\pm2.7$ \\
	&	$25.0-26.0$	&	35 &	$3.1\pm1.6$ &   $8.8\pm1.4$	&	$23.6\pm2.3$	&	$27.4\pm2.2$	&	$18.2\pm1.9$ \\
	&	$26.0-27.0$	&	30 &	$6.8\pm1.3$ &	$12.8\pm3.6$	&	$26.4\pm5.3$	&	$33.5\pm6.1$	&	$30.9\pm8.3$ \\
	&	$27.0-28.5$	&	8  &	$7.5\pm2.3$ &	$7.2\pm3.4$	&	$28.1\pm4.0$	&	$31.3\pm4.2$	&	$36.9\pm5.3$	 \\
		
		&				&		&				&				&					&					&	\\

RQQs	&	$21.0-23.0$	&	228 &	$7.4\pm0.6$ &	$9.7\pm0.7$	&	$20.9\pm1.1$	&	$19.4\pm0.7$	&	$11.8\pm0.6$ \\
		&	$23.0-23.5$	&	249	&	$5.3\pm0.8$	&	$8.6\pm1.3$	&	$20.4\pm2.1$	&	$21.1\pm1.3$	&	$13.7\pm0.8$\\
		&	$23.5-24.0$	&	378 &	$4.3\pm0.3$ &	$8.9\pm0.4$	&	$18.8\pm0.7$	&	$21.3\pm0.6$	&	$14.5\pm0.5$	\\
		&	$24.0-25.5$	&	152	 &	$4.5\pm0.7$	&	$7.6\pm1.2$	&	$19.1\pm1.8$	&	$22.5\pm1.2$	&	$16.9\pm1.1$ \\

\hline
\hline

Class	&	$\log_{10}(L_{\rm opt}/{\rm W})$ 	&	$N$ per	bin	&	\multicolumn{5}{c}{Mean flux density (mJy)}		\\
		&							&	  &	100 $\mu$m	&	160 $\mu$m	&	250 $\mu$m	&	350 $\mu$m	&	500 $\mu$m 	\\

\hline

RLQs	&	$37.3-38.5$	&	31 	&	$6.8\pm1.7$	&	$12.2\pm2.2$	&	$23.4\pm3.0$	&	$28.2\pm2.7$	&	$20.1\pm2.3$	\\
	&	$38.5-39.0$	&	32 	&	$7.2\pm1.5$	&	$9.9\pm3.7$	&	$19.5\pm2.1$	&	$25.5\pm1.8$	&	$18.0\pm1.8$	\\
	&	$39.0-40.3$	&	30 	&	$4.3\pm1.4$	&	$8.0\pm1.4$	&	$29.8\pm4.8$	&	$35.8\pm4.6$	&	$36.0\pm5.6$  \\

		&				&		&				&					&					&					&	\\

RQQs	&	$37.3-38.5$	&	301 &	$5.9\pm0.5$	&	$9.0\pm0.6$	&	$19.1\pm0.9$	&	$19.6\pm0.6$	&	$12.6\pm0.6$\\
	&	$38.5-39.0$	&	400 &	$5.1\pm0.5$	&	$8.2\pm0.8$		&	$18.7\pm1.2$	&	$20.6\pm0.8$	&	$13.5\pm0.6$	 \\
	&	$39.0-40.3$	&	306 &	$5.0\pm0.4$	&	$9.5\pm0.7$		&	$21.7\pm1.1$	&	$22.9\pm0.8$	&	$16.1\pm0.7$	 \\

\hline

\end{tabular}
\label{Table:mean_fluxes}
\end{center}
\end{table*}

\begin{table*}
\begin{center}

\caption{The K-S (left column) and M-W (right column) probabilities of radio-loud quasars flux densities being indistinguishable from radio-quiet quasars in redshift and optical luminosity bins at 100, 160, 250, 350 and 500 $\mu$m, respectively.}
\centering
\begin{tabular} {c c c c c c }

\hline 

$z$-range	&	\multicolumn{5}{c}{K-S/M-W probability (\%)}	\\
		&	 100 $\mu$m	&	160 $\mu$m	&		250 $\mu$m	&	350 $\mu$m	&	500 $\mu$m 			\\

\hline

$0.0-1.0$	&	41.0/22.3	&	10.5/6.7	&	82.2/15.0	&	0.9/4.2		&	0.1/0.3		\\
$1.0-1.5$	&	44.8/38.8	&	67.0/39.4	&	54.6/31.7	&	3.5/4.8		&	3.6/4.9		\\
$1.5-2.0$	&	88.9/30.8	&	56.0/35.2	&	96.4/29.2	&	57.5/36.4	&	39.9/7.6		\\		
$2.0-4.0$	&	80.5/39.4	&	14.4/4.1	& 18.7/3.3	&	64.6/15.8	&	67.9/26.7	\\
		
\hline
\hline

$\log_{10}(L_{\rm opt}/{\rm W})$ 	&	\multicolumn{5}{c}{K-S/M-W probability (\%)}	\\
					&	100 $\mu$m	&	160 $\mu$m	&		250 $\mu$m	&	350 $\mu$m	&	500 $\mu$m 		\\

\hline

$37.3-38.5$	&	55.7/37.1 	&	80.6/40.9	&	36.9/27.7 	&	3.8/2.9 	&	0.7/0.4 	\\
$38.5-39.0$	&	51.6/9.5	&	16.6/7.6	& 	93.2/42.0	&	20.8/8.8	&	12.4/3.0	\\
$39.0-40.3$	&	99.7/46.9	&	17.9/6.7	& 	32.6/4.1	&	70.9/13.4	&	71.2/37.4	\\

\hline

\end{tabular}
\label{Table:KS_fluxes}
\end{center}
\end{table*}

\subsection{Luminosity calculation}

To convert between measured FIR flux density at \emph{Herschel} wavelengths and total luminosity in the FIR band and to derive the dust temperature, we have to adopt a model for the FIR spectral energy distribution (SED). We use a single temperature grey-body fitting function \citep{Hildebrand1983} in which the thermal dust spectrum is approximated by: $F_{\nu}=\Omega Q_{\nu}B_{\nu}(T)$, where $B_{\nu}$ is the Planck function, $\Omega$ is the solid angle, $Q_{\nu}=Q_{0}({\nu} / {\nu}_{0})^{\beta}$ is the dust emissivity (with 1$\leq {\beta} \leq$2) and $T$ is the effective dust temperature. Since $T$ and $\beta$ are degenerate for sparsely sampled SEDs, following \cite{2010A&A...518L..10D} we have fixed the dust emissivity index to $\beta= 2.0$ and varied the temperature over the range $10<T(\rm K)<60$. The selection of the $\beta$ parameter has been made based on the $\chi^{2}$ value. Using a $\beta = 2.0$ instead of e.g. 1.5, the best-fitting model returns lower $\chi^{2}$ values for both of the populations. For each source we estimated the integrated FIR luminosity (8 -- 1000 $\mu$m) using the grey-body fitting with the best fit temperature. The dust temperature was obtained from the best fit model derived from minimization of the $\chi^{2}$ values. The uncertainty in the measurement was obtained by mapping the $\Delta \chi^{2}$ error ellipse. In addition to the integrated FIR luminosity we calculate the monochromatic FIR-luminosity at 250 $\mu$m, where the temperature-luminosity relation affects only the k-correction parameter, which is far less sensitive than the integrated FIR to the dust temperature \citep[e.g.][]{Jarvis2010,Hardcastle2013,Virdee2013}.

\section{Far-infrared properties}\label{sec:FIR}

In order to estimate the FIR properties of our samples based on the isothermal grey-body model, we use Levenberg-Marquardt $\chi^{2}$ minimization to find the best-fitting temperature and normalization value  for the grey-body model. The errors on the parameters were determined by mapping the $\Delta \chi ^{2}=2.3$ error ellipse, which corresponds to the 1$\sigma$ error for 2 parameters of freedom. 
For every source in our sample, we calculate the integrated FIR luminosity ($8-1000 ~\mu$m), the monochromatic luminosity at 250 $\mu$m and the isothermal dust mass using the 250-$\mu$m luminosity. The mass derived on the assumption of a single temperature for the dust, is given by: 
\begin{equation}
M_{\rm dust}=\frac{L_{250}}{4 \pi \kappa_{250} B(\nu_{250},T)}
\end{equation}
where $\kappa_{250}$ is the dust mass absorption coefficient, which \cite{Dunne2011} take to be $0.89~{\rm m^{2}~kg^{-1}}$ and $B(\nu, T)$ is the Planck function. K-corrections have been applied \footnote{The K-correction is given by: 
\\
$K=\left(\frac{\nu_{\rm obs}}{\nu_{\rm obs(1+z)}} \right)^{3+\beta}\frac{e^{(h\nu_{\rm obs(1+z)}/kT_{\rm iso})}-1}{e^{(h\nu_{\rm obs}/kT_{\rm iso})}-1} $, where $\nu_{\rm obs}$ is the observe frequency at 250$\mu$m, $\nu_{\rm obs(1+z)}$ is the rest-frame frequency and $T_{\rm iso}$ and $\beta$ are the temperature and emissivity index.}.

\subsection{Stacking}\label{sec:Stacking}

The majority of our sources are undetected at the 5$\sigma$ limit of the Phase 1 catalogue so, in order to calculate their properties we use two different stacking methods and we compare the results. The first method is based on a weighted stacking analysis which follows the method of \cite{Hardcastle2010}. We determine the luminosity for each source from the background-subtracted flux density, even if negative, on the grounds that this is the maximum-likelihood estimator of the true luminosity, and take the weighted mean of the parameter we are interested in within each bin. We use the same redshift and optical luminosity bins across the radio-loud and radio-quiet samples in order to facilitate comparisons. The luminosity is weighted using the errors calculated from $\Delta \chi^{2}=2.3$ and the errors on the stacked parameters are determined using the bootstrap method. The advantage of bootstrapping is that no assumption is made on the shape of the luminosity distribution. Tables~\ref{Table:weighted_mean} and \ref{Table:weighted_meanKS} show the weighted mean values of the estimated parameter within each bin for both populations and the K-S/M-W test probabilities of the individual measurements comparing the radio-loud and radio-quiet quasars in the same bins.

\begin{table*}
\caption{Estimated weighted-mean far-infrared properties using a single-component grey-body fitting, K-S and M-W probabilities that the estimations for the radio-loud quasars in redshift, radio luminosity and optical luminosity bins are drawn from the same population as radio-quiet quasars, as a function of quasars class and parameter.}
\centering

\begin{tabular} {c c c c c c}

\hline 

Class	&	$z$		&	\multicolumn{4}{c}{Weighted mean values}	\\
		&	range	&	$\log_{10}({\rm LFIR}/{\rm L_{\odot}})$		&	$T_{\rm iso}~({\rm K})$	 	&	$\log_{10}(M_{\rm dust}/{\rm M_{\odot}}) $ 	&	$\log_{10}(L_{250}/{\rm W~Hz^{-1}})$ 	\\

\hline

RLQs	&	$0.0-1.0$	&	$11.11\pm0.07$	&	$18.42\pm1.30$	&	$7.79\pm0.08$	&	$25.80\pm0.08$	\\
	&	$1.0-1.5$	&	$11.90\pm0.06$	&	$19.41\pm1.26$	&	$8.00\pm0.11$	&	$26.68\pm0.12$ \\
	&	$1.5-2.0$	&	$12.17\pm0.06$	&	$25.79\pm1.49$	&	$8.06\pm0.09$	&	$27.07\pm0.10$	\\
	&	$2.0-4.0$	&	$12.38\pm0.09$	&	$27.14\pm1.43$	&	$7.77\pm0.08$	&	$27.26\pm0.26$	\\

	&			&			&			&			&			\\	
		
RQQs	&	$0.0-1.0$	&	$11.23\pm0.17$	&	$22.48\pm0.35$	&	$7.91\pm0.02$	&	$25.96\pm0.04$		\\
	&	$1.0-1.5$	&	$11.97\pm0.12$	&	$26.28\pm0.40$	&	$8.08\pm0.02$	&	$27.01\pm0.05$		\\
	&	$1.5-2.0$	&	$12.22\pm0.11$	&	$26.35\pm0.36$	&	$8.15\pm0.02$	&	$27.28\pm0.03$		\\
	&	$2.0-4.0$	&	$12.68\pm0.04$	&	$30.29\pm0.46$	&	$8.33\pm0.03$	&	$28.15\pm0.08$		\\
		
\hline
\hline

Class	&	$\log_{10}(L_{\rm 1.4GHz}/{\rm W~Hz^{-1}})$	 	&	\multicolumn{4}{c}{Weighted mean values}			\\
	&			&	$\log_{10}({\rm LFIR}/{\rm L_{\odot}})$		&	$T_{\rm iso}~({\rm K})$	 	&	$\log_{10}(M_{\rm dust}/{\rm M_{\odot}}) $ 	&	$\log_{10}(L_{250}/{\rm W~Hz^{-1}})$ 	\\

\hline

RLQs	&	$23.0-25.0$	&	$11.52\pm0.20$	&	$19.95\pm1.54$	&	$7.83\pm0.10$	&	$25.85\pm0.38$		\\
	&	$25.0-26.0$	&	$11.95\pm0.06$	&	$24.25\pm1.01$	&	$7.94\pm0.05$	&	$26.70\pm0.14$		\\
	&	$26.0-27.0$	&	$12.06\pm0.09$	&	$25.82\pm1.18$	&	$7.99\pm0.13$	&	$26.79\pm0.37$		\\
	&	$27.0-28.5$	&	$12.33\pm0.19$	&	$27.18\pm1.17$	&	$8.10\pm0.09$	&	$27.08\pm0.23$		\\

	&				&					&					&					&						\\	

RQQs	&	$21.0-23.0$	&	$11.17\pm0.16$	&	$22.09\pm0.39$	&	$7.59\pm0.03$	&	$25.87\pm0.05$		\\
	&	$23.0-23.5$	&	$11.89\pm0.17$	&	$25.91\pm0.45$	&	$8.05\pm0.03$	&	$26.86\pm0.09$	   \\
	&	$23.5-24.0$	&	$12.16\pm0.10$	&	$26.56\pm0.33$	&	$8.10\pm0.02$	&	$27.21\pm0.03$		\\
	&	$24.0-25.5$	&	$12.74\pm0.04$	&	$31.91\pm0.51$	&	$8.38\pm0.03$	&	$28.21\pm0.09$		\\

\hline
\hline	

Class	&	$\log_{10}(L_{\rm opt}/{\rm W})$	 	&	\multicolumn{4}{c}{Weighted mean values}		\\
	&					&	$\log_{10}({\rm LFIR}/{\rm L_{\odot}})$		&	$T_{\rm iso}~({\rm K})$	 	&	$\log_{10}(M_{\rm dust}/{\rm M_{\odot}}) $ 	&	$\log_{10}(L_{250}/{\rm W~Hz^{-1}})$ 	\\

\hline

RLQs	&	$37.3-38.5$	&	$11.74\pm0.08$	&	$18.89\pm1.29$	&	$7.90\pm0.02$	&	$26.57\pm0.25$		\\
	&	$38.5-39.0$	&	$11.94\pm0.07$	&	$19.98\pm1.18$	&	$8.10\pm0.03$	&	$26.89\pm0.11$	    \\
	&	$39.0-40.3$	&	$12.32\pm0.10$	&	$27.05\pm1.06$	&	$8.15\pm0.02$	&	$27.31\pm0.26$		\\
		
	&			&			&			&			&				\\

RQQs	&	$37.3-38.5$	&	$11.36\pm0.02$	&	$21.43\pm0.33$	&	$7.67\pm0.02$	&	$26.22\pm0.07$		\\
	&	$38.5-39.0$	&	$12.00\pm0.08$	&	$25.94\pm0.33$	&	$8.01\pm0.03$	&	$27.04\pm0.05$		\\
	&	$39.0-40.3$	&	$12.53\pm0.02$	&	$29.16\pm0.34$	&	$8.31\pm0.02$	&	$27.92\pm0.08$		\\

\hline

\end{tabular}
\label{Table:weighted_mean}
\end{table*}

\begin{table*}
\caption{The K-S and M-W probabilities that the estimations for the radio-loud quasars in redshift, radio luminosity and optical luminosity bins are drawn from the same population as radio-quiet quasars.}
\centering
\begin{tabular} { c c c c c }

\hline 

$z$-range		&	\multicolumn{4}{c}{K-S/M-W probability (\%)}	\\
	&	LFIR 	&	$T_{\rm iso}$	&	 $M_{\rm dust}$	&	$L_{250}$		 		\\

\hline

$0.0-1.0$	&	11.9/9.5	&	25.0/29.8	&	93.3/29.8	&	21.8/38.7	\\
$1.0-1.5$	&	95.7/39.5	&	60.6/24.8	&	25.8/26.2	&	38.2/33.1	\\
$1.5-2.0$	&	15.0/4.2	&	74.4/22.4	&	79.1/17.6	&	11.9/8.1	\\
$2.0-4.0$	&	27.7/6.7	&	6.0/1.0		&	7.6/1.1	&	21.2/4.8	\\
		
\hline
\hline

$\log_{10}(L_{\rm opt}/{\rm W})$	 	&		\multicolumn{4}{c}{K-S/M-W probability (\%)}	\\
		&	LFIR 	&	$T_{\rm iso}$	&	 $M_{\rm dust}$	&	$L_{250}$			\\

\hline

$37.3-38.5$	&		4.1/0.6	&	9.1/21.6	&	32.9/12.8	&	2.1/4.6	\\
$38.5-39.0$	&		18.6/2.6	& 	1.0/1.0		&	35.6/6.3	& 	1.44/0.5\\
$39.0-40.3$	&		36.4/10.9	&	4.0/3.0	&	53.6/13.1	&	2.0/0.9	\\

\hline

\end{tabular}
\label{Table:weighted_meanKS}
\end{table*}


Using the weighted stacking analysis might bias our measurement to the brightest and hottest objects. In order to ensure that the FIR parameters from the weighted stacking method are reliable, we calculate, as an alternative, the mean temperatures for objects using the Maximum Likelihood Temperature method \citep[e.g.][]{Hardcastle2013}. As in the previous sections, we split the radio-loud and radio-quiet quasars into bins defined by their redshift, optical luminosity and radio luminosity. For each bin, we calculate the best fit temperature that gives the best $\chi^{2}$ fit to the observed fluxes of every quasar in the bin. In order to do this, we cycle through temperatures between 5 - 60 K allowing each quasar to vary and have a free normalization. For each temperature step, we calculate the total $\chi^{2}$. This result is a distribution from which we determine the temperature with the lowest total $\chi^{2}$. Errors in this fitted temperature are estimated by finding the range that gives $\Delta \chi^{2}=1$. Using the best-fitting temperature and normalizations for all the galaxies, we estimate the FIR luminosity, the 250-$\mu$m luminosity and the dust mass for each bin. The errors for each parameter are determined by bootstrapping. The results of this method are shown in Table \ref{Table:ML}. The advantages of this method are that all the sources in a given bin are used in the temperature estimation and the luminosities of the sources in bins are not automatically correlated. However, there are bins where the estimated mean temperature is significantly different from the individual temperature of each source, which could result in underestimation (or overestimation) of luminosities and dust masses.

\begin{table*}
\caption{Mean far-infrared parameters for each bin as they are estimated by the Maximum Likelihood (ML) stacking method. The best $\chi^{2}$ for each bin is also presented.}
\centering
\begin{tabular} {c c c c c c c }

\hline 

Class	&	$z$-range		&	\multicolumn{4}{c}{ML mean values}		&	$\chi^{2}$	\\
		&		&	$\log_{10}({\rm LFIR/L_{\odot}})$	&	$T_{\rm iso} ~({\rm K})$	&	$\log_{10}(M_{\rm dust} /{\rm M_{\odot}}) $	&	$\log_{10}(L_{250}/{\rm W~Hz^{-1}})$	&			\\

\hline

RLQs	&	$0.0-1.0$	&	$11.19\pm0.08$		&	$18.01^{+1.01}_{-0.72}$		&	$7.88\pm0.07$	&	$25.81\pm0.08$		&	1.11 	\\
		&	$1.0-1.5$	&	$11.81\pm0.07$		&	$21.72^{+1.61}_{-1.35}$		&	$8.26\pm0.09$	&	$26.70\pm0.08$		&	0.68	\\
		&	$1.5-2.0$	&	$12.04\pm0.05$		&	$25.76^{+1.32}_{-1.13}$		&	$8.10\pm0.08$	&	$27.00\pm0.07$		&	0.56	\\
		&	$2.0-4.0$	&	$12.42\pm0.08$		&	$27.22^{+1.54}_{-2.74}$		&	$7.97\pm0.08$	&	$27.19\pm0.08$		&	0.14	\\

		&				&						&								&					&					&		\\
		
RQQs	&	$0.0-1.0$	&	$11.11\pm0.02$		&	$21.27^{+0.41}_{-0.38}$		&	$7.78\pm0.02$	&	$25.99\pm0.03$		&	0.69	\\
		&	$1.0-1.5$	&	$11.86\pm0.03$		&	$24.19^{+0.53}_{-0.42}$		&	$8.09\pm0.03$	&	$26.88\pm0.03$		&	0.50	\\
		&	$1.5-2.0$	&	$12.21\pm0.02$		&	$27.24^{+0.77}_{-0.43}$		&	$8.21\pm0.02$	&	$27.24\pm0.02$		&	0.46	\\
		&	$2.0-4.0$	&	$12.56\pm0.17$		&	$30.26^{+1.20}_{-0.84}$		&	$8.27\pm0.04$	&	$27.88\pm0.04$		&	0.67	\\

\hline
\hline

Class	&	$\log_{10}(L_{\rm 1.4GHz}/{\rm W~Hz^{-1}})$ 		&	\multicolumn{4}{c}{ML mean values}		&	$\chi^{2}$	\\
		&		&	$\log_{10}({\rm LFIR/L_{\odot}})$	&	$T_{\rm iso} ~({\rm K})$	&	$\log_{10}(M_{\rm dust} / {\rm M_{\odot}}) $	&	$\log_{10}(L_{250}/{\rm W~Hz^{-1}})$	&		\\

\hline

RLQs	&	$23.0-25.0$	&	$11.19\pm0.13$	&	$17.79^{+2.75}_{-0.46}$	&	$7.81\pm0.12$	&	$25.77\pm0.16$	&	1.06	\\
		&	$25.0-26.0$	&	$11.93\pm0.06$	&	$22.25^{+1.29}_{-1.50}$	&	$8.05\pm0.05$	&	$26.62\pm0.05$	&	0.60	\\
		&	$26.0-27.0$	&	$12.09\pm0.07$	&	$25.03^{+1.98}_{-1.64}$	&	$8.10\pm0.10$	&	$26.91\pm0.07$	&	0.47	\\
		&	$27.0-28.5$	&	$12.55\pm0.15$	&	$30.26^{+1.26}_{-3.16}$	&	$8.15\pm0.10$	&	$27.19\pm0.08$	&	0.23	\\

		&				&						&							&					&					&		\\
		
RQQs	&	$21.0-23.0$	&	$11.03\pm0.03$	&	$19.75^{+0.43}_{-0.32}$	&	$7.79\pm0.02$	&	$25.84\pm0.03$	&	0.69	\\
		&	$23.0-23.5$	&	$11.76\pm0.04$	&	$22.58^{+1.75}_{-0.42}$	&	$8.11\pm0.04$	&	$26.61\pm0.05$	&	0.50	\\
		&	$23.5-24.0$	&	$12.16\pm0.01$	&	$27.55^{+0.58}_{-0.42}$	&	$8.17\pm0.02$	&	$27.19\pm0.02$	&	0.48	\\
		&	$24.0-25.5$	&	$12.68\pm0.16$	&	$30.57^{+1.37}_{-0.73}$	&	$8.20\pm0.04$	&	$27.82\pm0.03$	&	0.70	\\

\hline
\hline	

Class	&	$\log_{10}(L_{\rm opt}/{\rm W})$ 		&	\multicolumn{4}{c}{ML mean values}		&	$\chi^{2}$	\\
		&		&	$\log_{10}({\rm LFIR/L_{\odot}})$	&	$T_{\rm iso} ~({\rm K})$	&	$\log_{10}(M_{\rm dust} /{\rm M_{\odot}}) $	&	$\log_{10}( L_{250}/{\rm W~Hz^{-1}})$	&	\\

\hline

RLQs	&	$37.3-38.5$	&	$11.62\pm0.10$	&	$17.25^{+3.12}_{-1.20}$	&	$8.00\pm0.09$	&	$25.95\pm0.11$	&	1.96	\\
		&	$38.5-39.0$	&	$11.93\pm0.06$	&	$21.25^{+1.28}_{-0.96}$	&	$8.18\pm0.06$	&	$26.50\pm0.05$	&	1.20	\\
		&	$39.0-40.3$	&	$12.36\pm0.07$	&	$26.52^{+2.16}_{-1.21}$	&	$8.09\pm0.08$	&	$27.12\pm0.06$	&	1.56	\\

		&				&						&							&					&					&		\\

RQQs	&	$37.3-38.5$	&	$11.22\pm0.03$	&	$19.22^{+1.25}_{-0.75}$	&	$7.84\pm0.02$	&	$25.81\pm0.03$	&	0.71	\\
		&	$38.5-39.0$	&	$11.92\pm0.02$	&	$25.28^{+1.62}_{-0.35}$	&	$8.07\pm0.03$	&	$26.89\pm0.02$	&	0.44	\\
		&	$39.0-40.3$	&	$12.36\pm0.02$	&	$28.06^{+1.83}_{-1.72}$	&	$8.20\pm0.02$	&	$27.33\pm0.03$	&	0.45	\\
\hline

\end{tabular}
\label{Table:ML}
\end{table*}

In general terms, the two methods are in good agreement with some exceptions in the case of `sensitive' parameters related to temperature. Specifically, it seems that we get larger differences in bins where the objects span a greater range in temperature. In these cases, the weighted mean method is dominated by the hotter objects returning higher luminosities. Despite the differences in temperature between the methods, we see that the monochromatic luminosities are broadly consistent in both methods implying that the temperature-luminosity correlation does not have a significant effect on the inferred monochromatic luminosities. In contrast, FIR luminosity and dust masses seem to be affected when hot objects are present. Despite the differences we get in some cases, both methods show that radio-loud quasars have systematically lower dust temperature than radio-quiet quasars. Regarding their luminosities, and especially the 250-$\mu$m luminosity which seems to be a safer choice as it is less affected by temperature, they tend to be comparable for most of the bins but not at lower optical luminosities (and/or redshifts) where an excess in the case of radio-loud quasars is found. 

\begin{figure*}
\begin{center}
\includegraphics[scale=0.495]{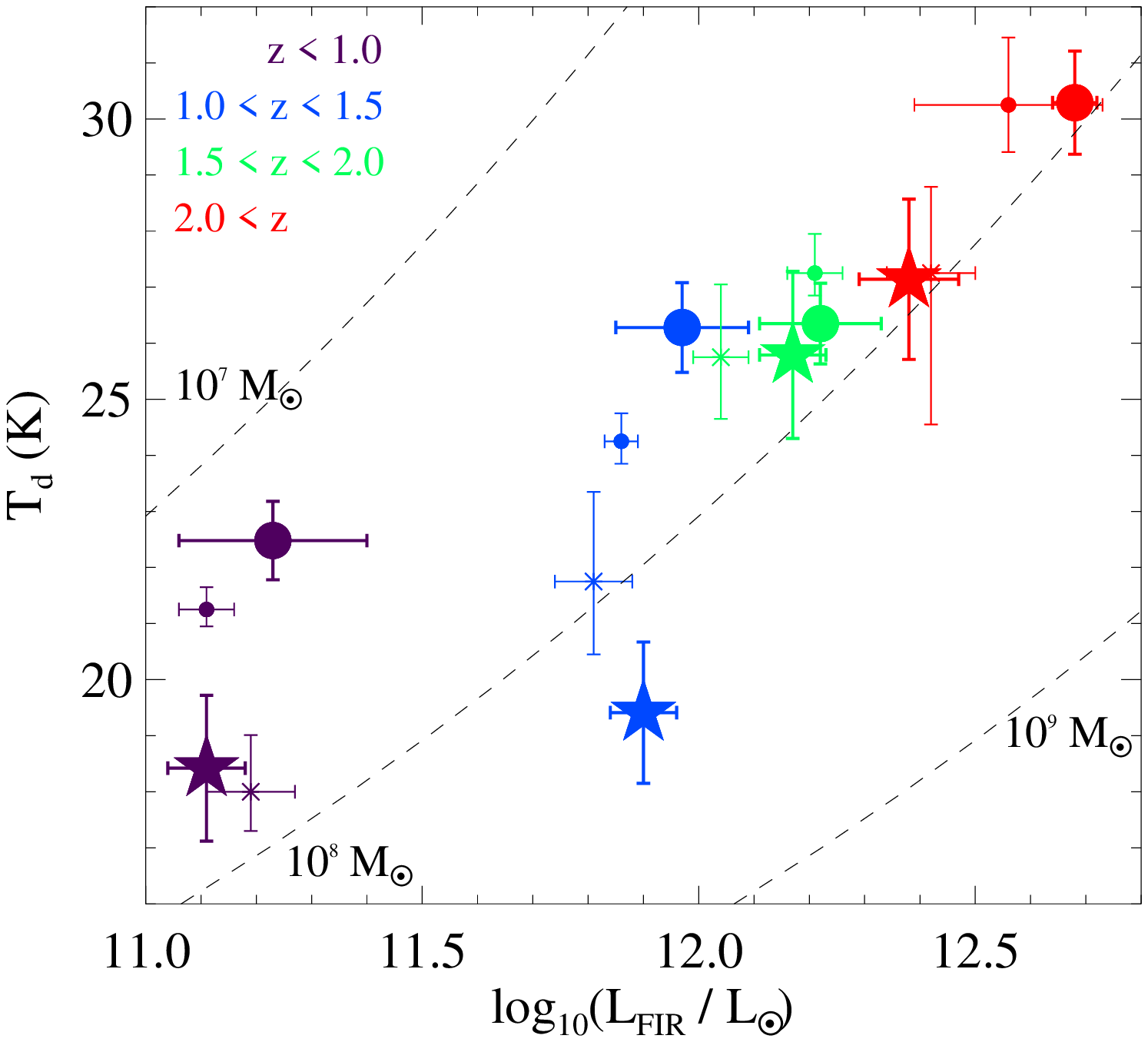}
\includegraphics[scale=0.495]{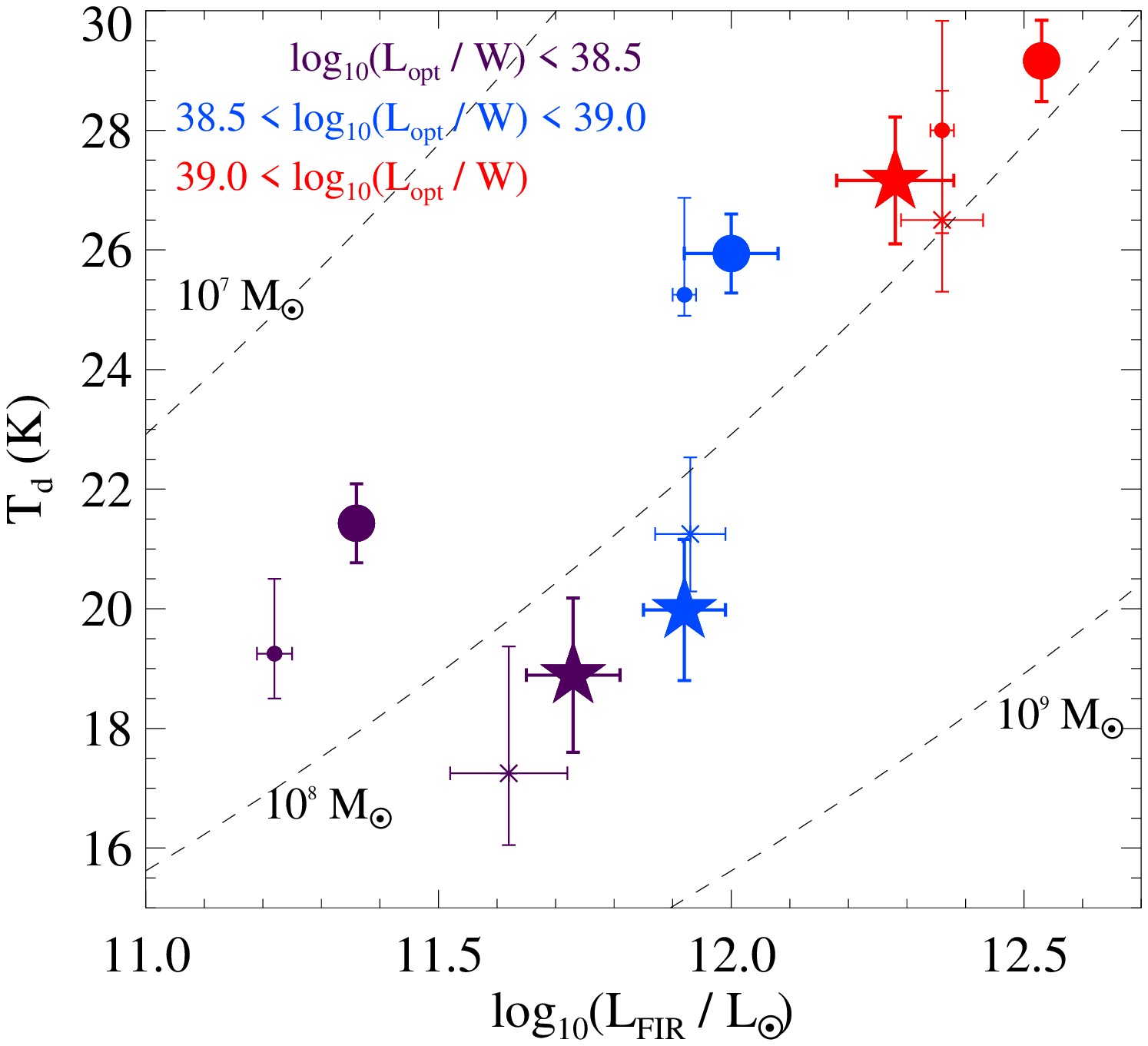}
\end{center}
\caption{FIR luminosity versus dust temperature when the two populations are divided into redshift (left) and optical luminosity (right) bins. The rainbow colour-code represents the redshift/optical lumimosity bin values, purple for lower and red for higher values respectively. The radio-loud quasars are represented by stars while the radio-quiet quasars are shown as circles. The large symbols show the estimates based on the weighted mean method while the small symbols show the estimates based on the maximum likelihood stacking. The black lines correspond to the dust mass estimates based on the LFIR - $T_{\rm dust}$ relation (LFIR$\propto \kappa_{0}M_{\rm dust} T_{\rm dust}^{4+\beta}$), assuming $\beta=2.0$, for dust masses of $10^{7}$, $10^{8}$ and $10^{9}~{\rm M_{\odot}}$.}
\label{fig:sum_up}
\end{figure*}

In order to study the FIR properties (e.g. FIR luminosity, dust temperature, dust mass) for the two populations as a function of redshift and optical luminosity we present in Fig.~\ref{fig:sum_up} the mean dust temperature as a function of the mean FIR luminosity. The two populations have been divided into redshift (left) and optical luminosity (right) bins which are represented by a rainbow colour-code with purple colour for lower and red colour for higher values. For each bin the weighted mean and the ML mean values are presented.

\subsection{FIR luminosity}
With respect to the redshift bins (see Fig.~\ref{fig:sum_up} left), the two populations have the same mean FIR luminosities within their errors for each bin. The largest difference between the mean FIR luminosities of the two populations is observed at the highest redshift bin ($z>2.0$; red colour) with radio-quiet quasars having the higher FIR luminosity. However, this difference could be an effect of the calculation of the mean values as both methods do not return significant excess for the radio-quiet population (small versus large red symbols). To summarize, the mean FIR luminosities of the radio-loud and radio-quiet quasars show no significant differences when the two population are split into redshift bins.
In contrast, when we divide the two populations into optical luminosity bins (see Fig.~\ref{fig:sum_up} right), there is a clear excess of FIR luminosity in lower-luminosity bins for the case of radio-loud quasars ($\log_{10}(L_{\rm opt}/{\rm W})<38.5$; purple colour). The fact that both methods show the same significant excess indicates that the observed differences between the two populations are not a result of the calculation methods. At intermediate optical luminosities ($38.5<\log_{10}(L_{\rm opt}/{\rm W})<39.0$; blue colour) both of the populations have consistent mean FIR luminosity values. At the highest optical luminosity bin ($\log_{10}(L_{\rm opt}/{\rm W})>39.0$; red colour) we have the same picture as at the highest redshift bin; a possible FIR luminosity excess for the radio-quiet quasars which, however, is not supported by both of the methods.

\subsection{Dust temperature and mass}
Our results reported in Fig.~\ref{fig:sum_up} and Tables~\ref{Table:weighted_mean},~\ref{Table:ML} show that there is a general trend that the radio-loud quasars have lower dust temperature than radio-quiet quasars, at least at lower redshift and optical luminosity bins. This difference reaches $\sim 5$ K in some bins. At higher redshift and optical luminosity bins both of the populations have the same mean dust temperatures within their errors. 

The mean values of the estimated dust mass based on both calculation methods show that radio-loud quasars have almost a constant mean dust mass over the whole redshift and optical luminosity range. In the case of radio-quiet quasars, the mean dust masses decrease at lower redshift/optical luminosity bins. Comparing the results for the two populations, it seems that radio-loud quasars have higher dust masses at lower luminosity bins while at higher luminosities both of the populations have similar mean values. Dust masses must be interpreted with care as they could be biased by the stacking analysis towards the brightest and hottest objects. The excess in dust mass, in the case of radio-loud quasars which are the class with the lower dust temperature, could be required in order to be detectable at a level that allows a temperature to be fitted.


\subsection{250-$\mu$m luminosity} \label{sec:250luminosity}

\begin{figure}
\centering
\includegraphics[scale=0.4]{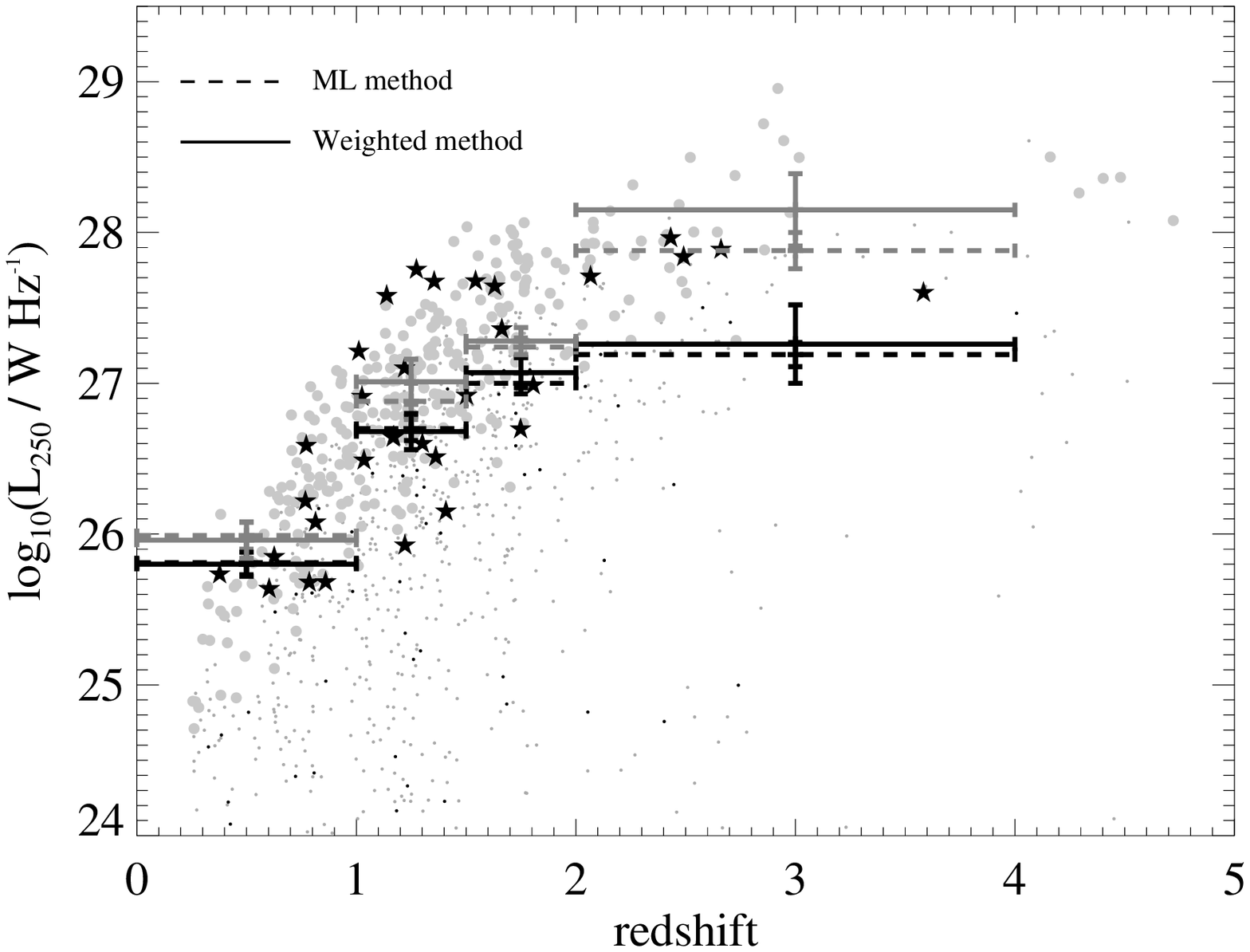}
\includegraphics[scale=0.4]{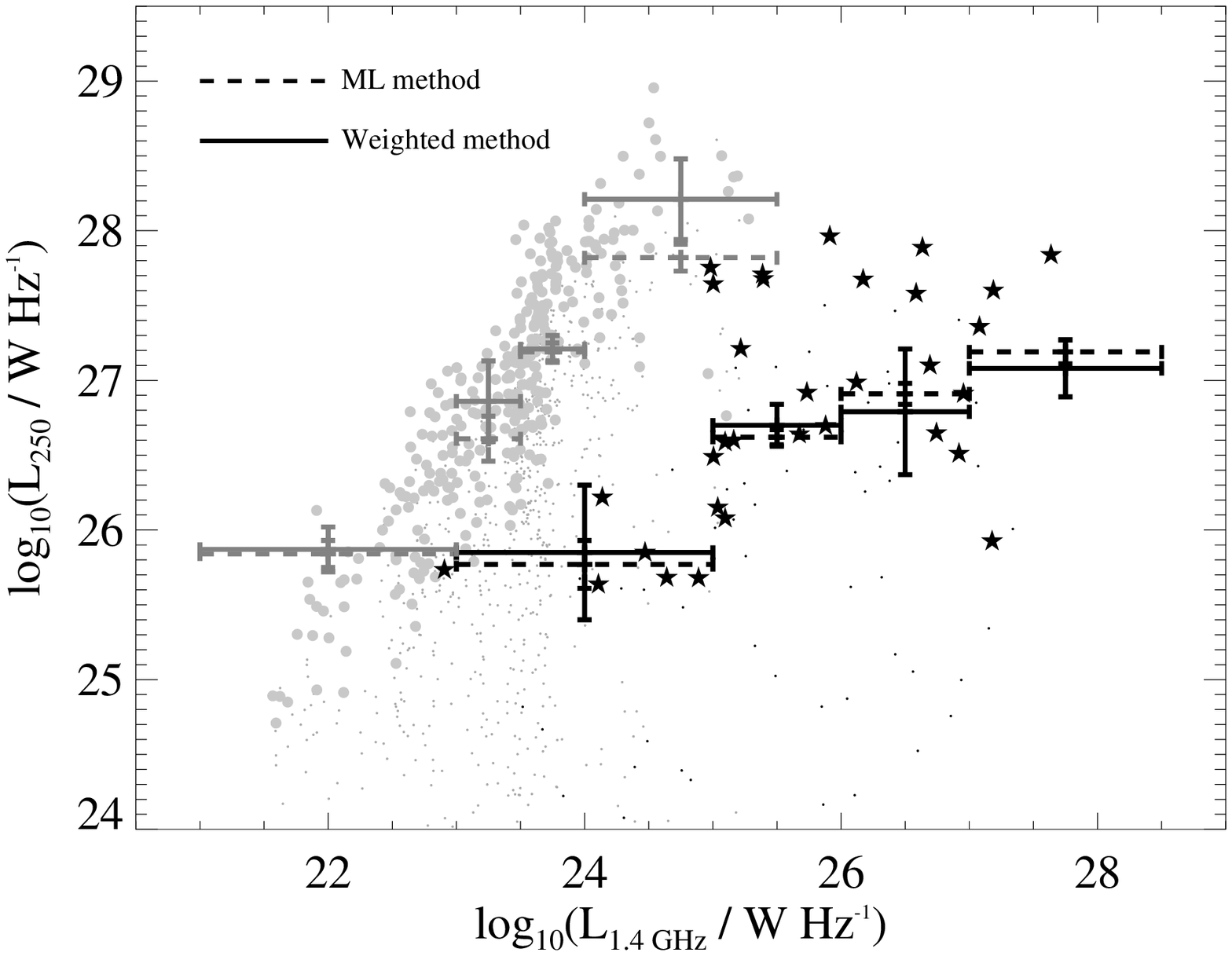}
\includegraphics[scale=0.4]{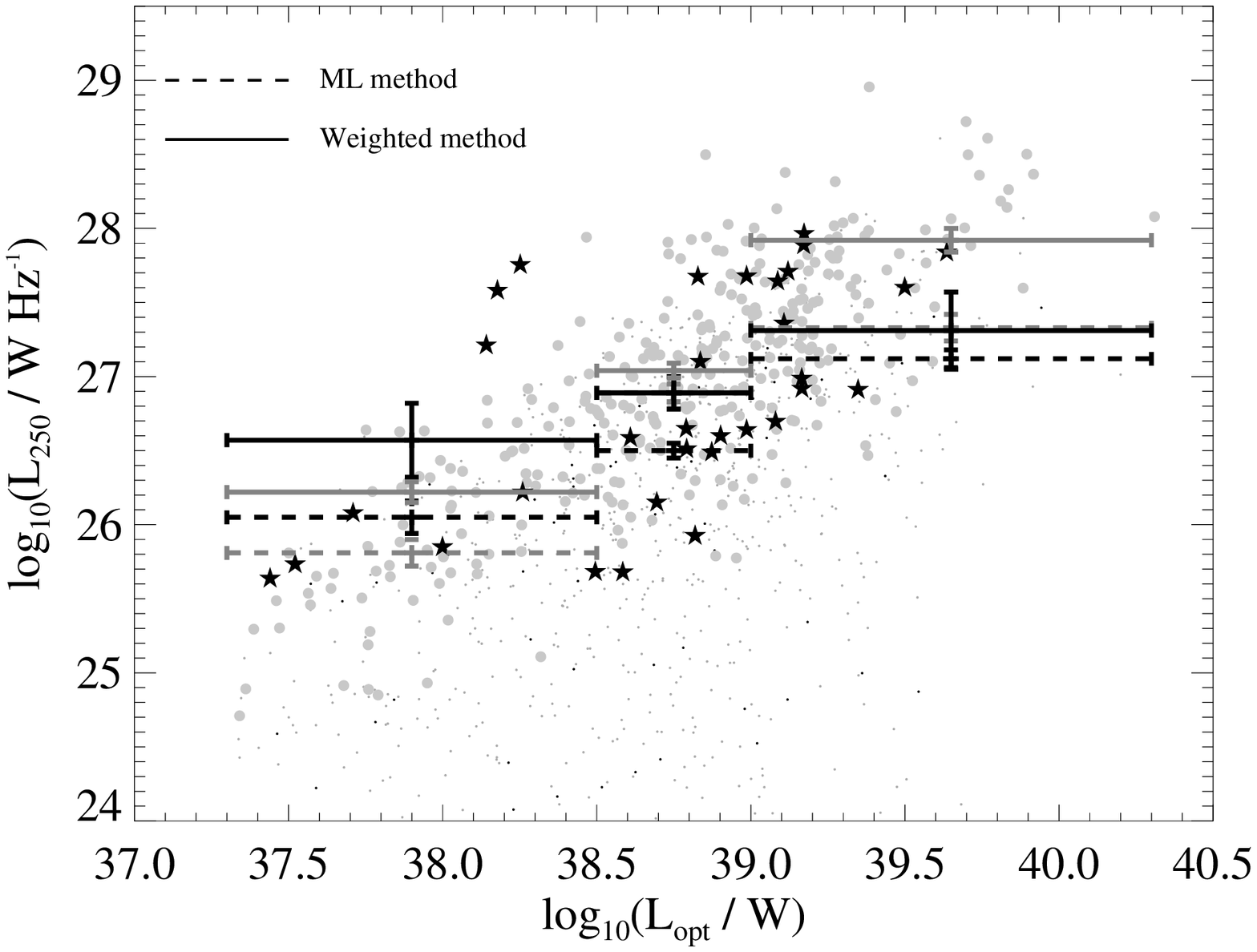}
\caption{Correlation between infrared luminosity at 250 $\mu$m as a function of redshift (top), radio luminosity (middle) and optical luminosity (bottom). Individual measurements for radio-loud (black stars) and radio-quiet (grey circles) quasars detected at 250 $\mu$m at the $3\sigma$ level are also included. Dots represent the entire samples. Error bars with solid lines illustrate the stacked luminosities calculated using the weighted method. Luminosities calculated via the Maximum Likelihood method are dashed line error bars. The errors have the same colour as the population that they represent.}
\label{fig:comparison}
\end{figure}

In this section we present the stacked monochromatic luminosity at 250 $\mu$m for both stacking methods and populations as a function of redshift, radio luminosity and optical luminosity (Fig.~\ref{fig:comparison}).
The luminosities calculated using the weighted stack method are shown by solid error bars while the luminosities calculated via the Maximum Likelihood method are shown by the dashed error bars. Both methods show a good level of agreement within their 1$\sigma$ error. The cases with the larger disagreement are those where strong outliers are found within the bin (unusually hot or cold sources in comparison with the rest of the population). Based on these plots, we see that the Maximum Likelihood Temperature method is more sensitive to outliers. We therefore argue that the weighted stacking method is sufficiently accurate to calculate the stacked rest-frame monochromatic luminosity at 250 $\mu$m. For clarity, we do not show the stacks generated by the Maximum Likelihood Temperature method in the subsequent sections, although consistency checks were performed throughout the analysis. 

As we see in Fig.~\ref{fig:comparison}, 250-$\mu$m luminosity is correlated with radio luminosity for both populations. However, the question is whether radio activity induces star formation, leading to FIR emission. Redshift will affect the correlation between the two luminosities so, as a first way to measure the strength of correlation between FIR luminosity and radio luminosity we use partial-correlation analysis \citep{Akritas1996}, which allows us to determine the correlation between the two parameters while accounting for the effects of redshift. For our analysis, we avoid bias against FIR weak sources by adding undetected sources (`censored' sample) to the detected sample. For this reason, in order to measure the partial correlations we use the FORTRAN program CENS-TAU, available from the Penn State Center for Astrostatistics\footnote{Available at\\ http://www.astrostatistics.psu.edu/statcodes/censtau.}, taking `censored' data into account as upper limits using the methodology presented in \cite{Akritas1996}.

The partial-correlation shows that radio luminosity is significantly correlated with 250-$\mu$m luminosity in the case of radio-loud quasars with a partial-correlation of $\tau=0.17$. The null hypothesis of zero partial correlation is rejected at the $3\sigma$ level. In the case of radio-quiet quasars we found that the correlation is not statistically significant with $\tau=0.06$ and a probability under the null hypothesis $p=0.11$. The results are almost the same even when we compare the integrated FIR luminosity to the radio luminosity but even more significant for the case of radio-loud quasars with the null hypothesis of no correlation rejected at higher than the $4\sigma$ level. Despite the results found for both the populations as a total, the different trends which we found for low ($\log_{10}(L_{\rm opt}/{\rm W})\leq38.5$) and high ($\log_{10}(L_{\rm opt}/{\rm W})>38.5$) optical luminosities lead us to investigate the correlations also for these sub-samples. In the case of radio-loud quasars, the significant correlation between radio luminosity and 250-$\mu$m luminosity remains only for the low optical luminosity bin with $\tau=0.12$ ($p<0.001$; the probability of no correlation) while for the high luminosity bin no significant correlation is found ($\tau=0.04$ and $p\simeq0.29$). In contrast, for radio-quiet quasars no correlation is again found for either low ($\tau=0.02$ and $p\simeq0.26$) or high ($\tau=0.03$ and $p\simeq0.36$) optical luminosity bins. Similar trends are also obtained when we compare the FIR luminosity with the radio luminosity for the two populations at lower and higher optical luminosities. In terms of radio-quiet quasars, all sources with $\log_{10}(L_{250}/{\rm W~Hz^{-1}})\geq 27.0$ are associated with optical luminosities above the threshold at which the dichotomy is found. At this level of 250-$\mu$m luminosity it seems that all correlations with optical luminosity, radio luminosity and, possibly, also with redshift, tend to disappear.  Regarding the correlation between $L_{250}$ and radio luminosity, a significant number of radio-quiet quasars with $\log_{10}(L_{250}/{\rm W~Hz^{-1}})\geq 27.0$ have radio-luminosity higher than $10^{24}~{\rm W/Hz}$, a limit often used for the distinction between radio-loud and radio-quiet population.

\subsection{Star-formation rate}

For the calculation of the star-formation rate (SFR) the FIR luminosity is required. As we discussed in Section~\ref{sec:Stacking} the FIR luminosity seems to be more sensitive to temperature dispersion compared to the 250-$\mu$m luminosity. In this case, the SFR estimation could be strongly affected by the dust temperature. On the other hand, the rest-frame monochromatic luminosity at 250 $\mu$m minimises the dispersion in our calculations and small differences are found, within their errors, between the two methods (weighted and maximum likelihood temperature). In addition, the FIR luminosity, as described using the two-temperature model, could be affected by a strong cold component. However, our results show that both FIR luminosity and $L_{250}$ are dominated by the warm component. For these reasons we prefer to use the warm dust component as a tracer of the current star formation, whose mass and luminosity are primarily an indicator of the star-formation rate \citep{Dunne2011,Smith2012a}.

In order to investigate how strongly and in which cases the warm-component FIR luminosity is affected by the temperature, we compare the warm-component 250-$\mu$m luminosity to the warm-component FIR luminosity as they were estimated using the two-temperature model. For both of the populations we found the same linear correlation, within the errors, between the warm 250-$\mu$m and integrated FIR luminosities. The linear regression between the warm 250-$\mu$m luminosity and the warm FIR luminosity is found with the ordinary least squares (OLS) bisector \citep{Isobe1990} fit being ${\rm LFIR_{RL}}\propto10^{0.66\pm0.01}L_{\rm 250RL}$; ${\rm LFIR_{RQ}}\propto10^{0.63\pm0.03}L_{\rm 250RQ}$ for radio-loud and radio-quiet quasars respectively. 
The same trends for both of the populations show that as long as we investigate only the differences of SFR between them, the selection of either the $L_{250}$ or the integrated LFIR as indicators of star formation would not affect our results or, at least, the effect should be the same for both populations.

The calculation of the SFR was performed using the equation by \cite{Kennicutt1998}:
\begin{equation}
\centering
{\rm SFR(M_{\odot}~yr^{-1})=4.5\times 10^{-44}LFIR~(erg~s^{-1})},
\end{equation}
which assumes a Salpeter IMF in the mass range $0.1-100 ~{\rm M_{\odot}}$, continuous starbursts of age 10 - 100 Myr, and requires the integrated IR luminosity over the range 8 - 1000 $\mu$m. 

\begin{figure*}
\centering
\includegraphics[scale=0.43]{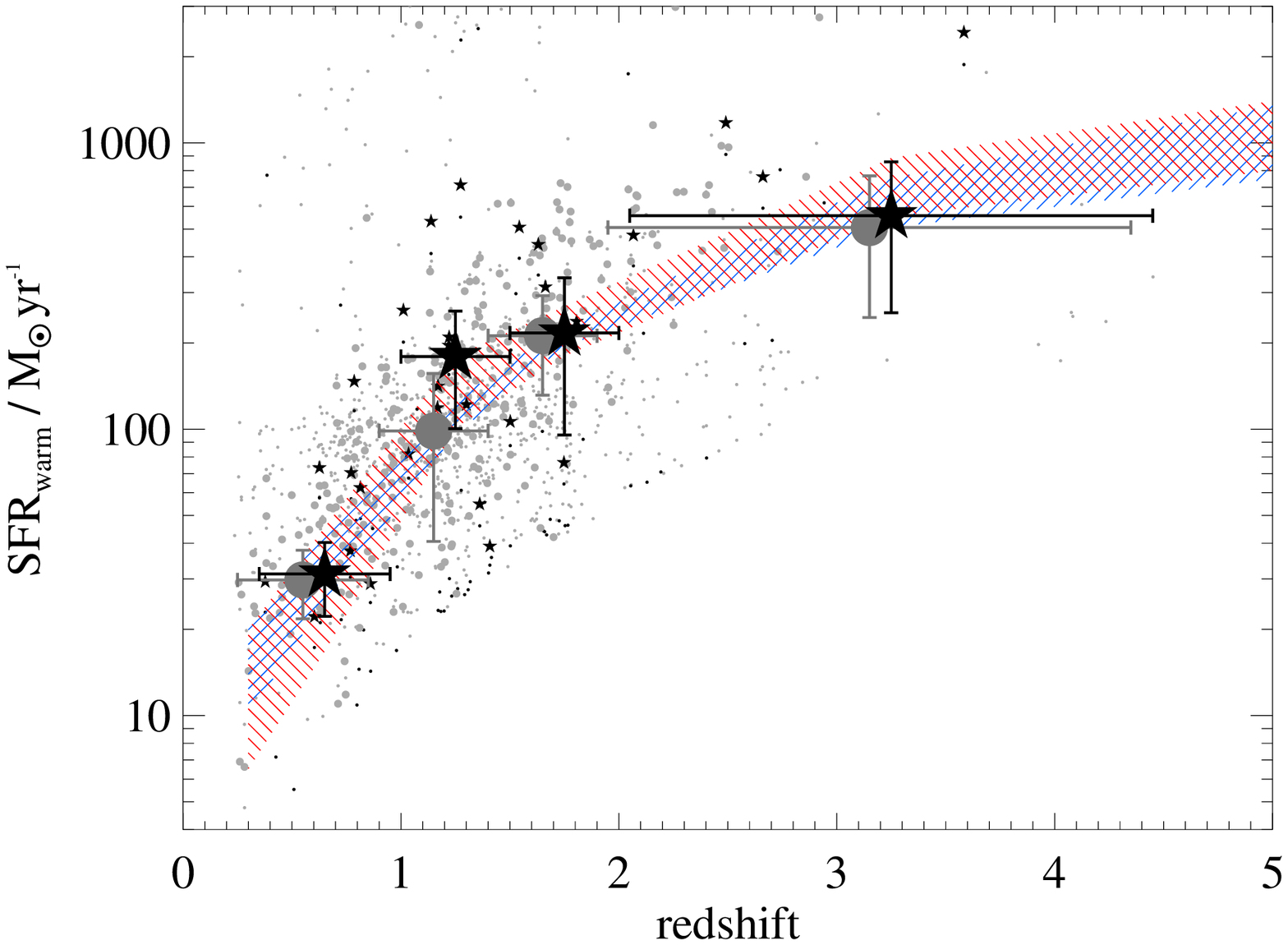}
\includegraphics[scale=0.43]{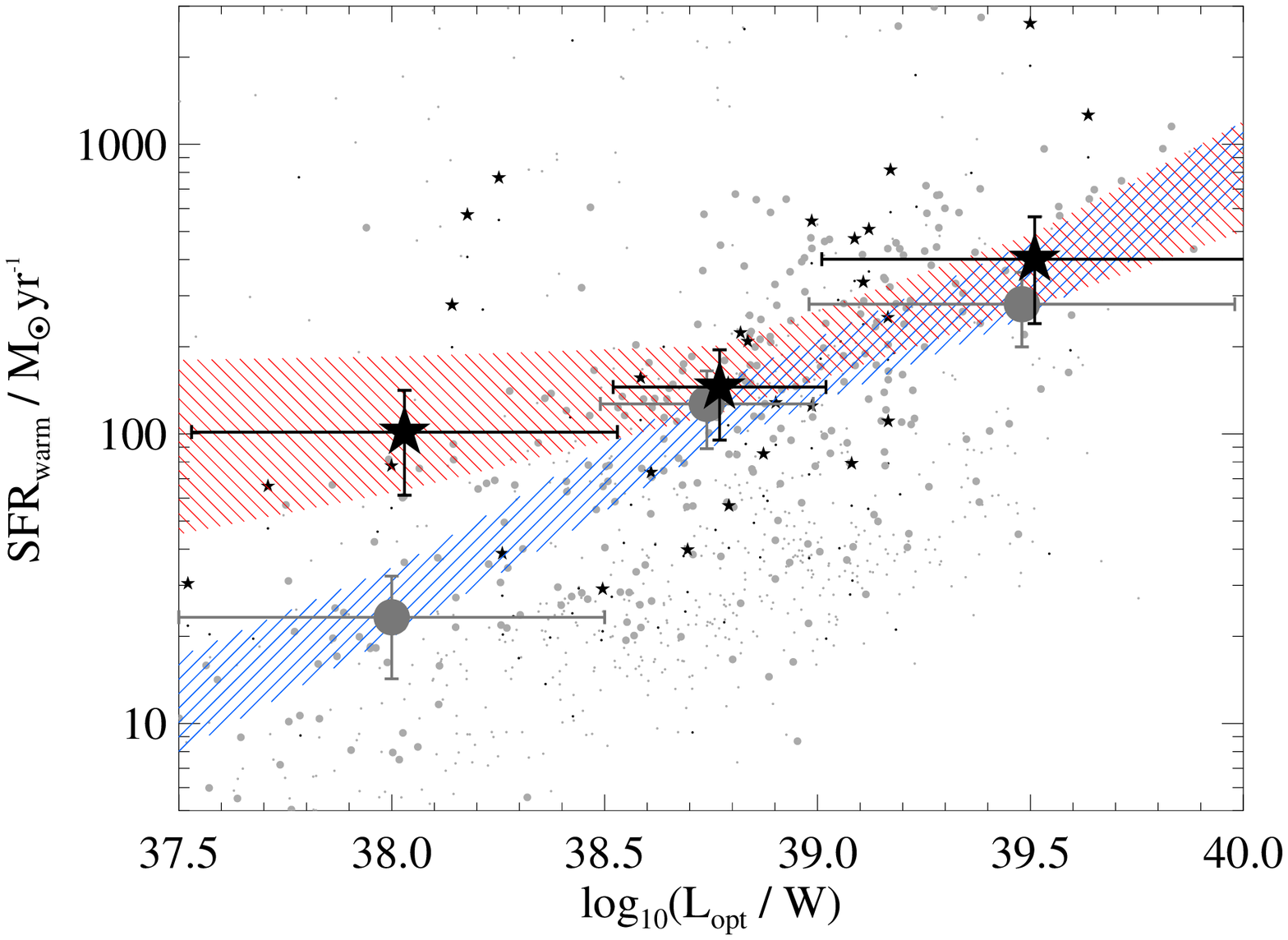}
\caption{Weighted mean star formation rates, $\left\langle {\rm SFR}\right\rangle $, as a function of redshift (left) and optical luminosity (right). The dots represent the entire sample. Small black stars represent the radio-loud quasars detected at 250 $\mu$m at the $3\sigma$ level. Small grey cycles represent the radio-quiet quasars detected at 250 $\mu$m at the $3\sigma$ level. The same but larger symbols for each population represent the weighted mean values based on the $T_{c}=15$~K, $T_{w}=35$~K two-temperature fitting model. The dashed regions (red for radio-loud and blue for radio-quiet quasars) show the range of the weighted mean values based on the $\pm 5$ K two-temperature fitting model regarding to the initial ($T_{c}=15$~K and $T_{w}=35$~K) choice of temperatures. In the left figure the large grey circles have been slightly left-shifted for clarity.}
\label{fig:SFR}
\end{figure*}

Fig.~\ref{fig:SFR} shows the weighted mean star-formation rates, $\left\langle {\rm SFR}\right\rangle $, derived from the warm-component FIR luminosities, as a function of optical luminosity and redshift, for radio-loud and radio-quiet quasars. We split the samples into 4 redshift and 3 optical luminosity bins trying to keep the same number of objects within each bin for each population and determined the SFR as described in Section \ref{2D_fitting}. The larger symbols represent the weighted mean SFR in each bin based on the $T_{c}=15$~K and $T_{w}=35$~K temperature fittings. Additionally, a dashed area is used to represent the mean values based on the different temperature pairs within $\pm 5$ K of the original temperatures. Taking into account the errors of the original mean values, it seems that the selection of the temperatures would not strongly affect our results as in most of the cases the errors are larger that the estimated differences between the different temperature models. Comparing the $\left\langle {\rm SFR}\right\rangle $ for the two populations as a function of redshift, no difference is found. Both radio-loud and radio-quiet quasars seem to have the same $\left\langle {\rm SFR}\right\rangle $ within their errors in each bin. Even if we take into consideration any possible combination of different temperature pairs, we would not observe any particular differences. On the other hand, comparing the $\left\langle {\rm SFR}\right\rangle $ as a function of optical luminosity, a significant excess is found in the case of radio-loud quasars for $\log_{10}(L_{\rm opt}/{\rm W})\leq38.5$. This difference remains significant even if we assume that the two populations have different dust temperatures. For $\log_{10}(L_{\rm opt}/{\rm W})>38.5$ both populations tend to have the same star-formation rate within their errors. Another interesting point is the presence of a possible break at $\log_{10}(L_{\rm opt}/{\rm W})\sim 38.5$ in the case of radio-loud quasars while radio-quiet quasars' data points could be easily described by a linear function.

\section{Discussion} \label{sec:discussion}

The results of the previous sections show that radio-loud quasars tend to have different FIR properties from a matched sample in redshift and optical luminosity of radio-quiet quasars. These differences lead to an excess of star-formation for the radio-loud population but are only significant in the case of low optical luminosity radio-loud and radio-quiet quasars.

Studying the FIR properties of an AGN population is usually a difficult task as possible contamination could affect the results. However, in this paper, we are mainly interested in studying the differences between the two populations instead of examining the exact properties for each one. In the case of our sample there are two main sources of contamination a) the warm dusty torus emission and b) the synchrotron emission of the powerful jets and lobes in the case of radio-loud quasars. In order to overcome these problems we followed two methods, one for each case.  We try to remove the problem of the warm dusty torus emission by matching our populations in redshift and optical luminosity. In this way, although we expect that FIR emission is largely uncontaminated by the AGN \citep[e.g][]{Haas2003,Hatziminaoglou2010}, any possible contamination would be the same for both populations. Different evolutionary models for the two populations could be also a possibility for different AGN contamination in the case of more evolved AGN, in which the BH gets closer to its final mass.  However, this could not affect our results as optical luminosity is a good tracer of the median accretion rate onto the central black hole and the Eddington ratio distribution is expected to be similar for the two populations at least at lower redshifts ($z<2.0$) and/or optical luminosity \citep[e.g.][]{Shankar2010} with both types of quasars being likely powered by similar physical mechanisms.

For the case of synchotron contamination, we estimated an upper limit on the possible contamination at FIR bands (see Appendix A). Based on these estimations, we either rejected contaminated objects from our sample or subtracted the synchrotron emission. Using these methods we consider our results to be unaffected by possible synchrotron contamination effects. 

\subsection{Star-formation excess}

Although the initial formation mechanisms of supermassive black holes remain largely unknown, the notion of seed black holes that form primordially and grow into a distribution of black hole masses has been around for four decades \citep[e.g.][]{Carr1974,Silk1998}. The mass distribution would necessarily be governed, at least partially, by the density of the surrounding gas; the most massive black holes would then form in regions of the highest gas density, and it will be in these sites where we observe high-redshift radio galaxies and radio-loud quasars. The highly relativistic, supersonic jets that power into the surrounding medium are able to trigger star formation along cocoons surrounding the jets \citep[e.g.][]{Bicknell2000,Fragile2004}. This model provides the means of orchestrating star formation over tens of kiloparsecs on light crossing timescales. This process has been invoked to explain the radio-optical alignment effect at high redshift \citep{Rees1989}. More recent, \cite{Drouart2014} suggested that radio galaxies have higher mean specific star formation rates (sSFR) than typical star-forming galaxies with the same black hole mass at least at higher redshifts, $z \leq 2.5$. 

Here we explore the link between radio AGN emission and star formation. Assuming that FIR luminosity is a good tracer of star formation, our results show a strong positive correlation between radio and FIR luminosity, independent of redshift, for radio-loud quasars (see Section~\ref{sec:250luminosity}). In contrast, no such correlation was found for radio-quiet quasars. Our results support the idea of a strong alignment between dust and jets from supermassive black holes. Powerful radio jets may increase the star-formation activity by compressing the intergalactic medium \citep[e.g.][]{SilkNusser10}, resulting in the observed star-formation excess we found for the radio-loud quasars. 

However, our results are not uniform over all the optical luminosity range of our sample. Radio-loud quasars seem to have higher star-formation rates (and FIR luminosities) than radio-quiet quasars only at lower optical luminosities. Specifically, we find that star-formation shows a possible break around to $\log_{10}(L_{\rm opt}/{\rm W})\approx 38.5$ in the case of radio-loud quasars. For lower optical luminosities, radio-loud quasars have higher star-formation than radio-quiet, while for higher optical luminosities both populations tend to have comparable $\left\langle {\rm SFR}\right\rangle $ within their errors. The same results were found no matter which method we used to estimate the FIR luminosity. This difference between the two populations could be an effect either of redshift or of AGN activity, as the optical luminosity is affected by both of these parameters. However, both populations seem to have the same FIR luminosity distribution over all redshifts within their errors. As the star-formation excess is not observed in the case of redshift distribution we deduce that the AGN activity is the main reason of this difference. Although we have found no strong evidence of star-formation suppression due to the radio activity at any redshift there are some hints like the decrease of the mean FIR flux densities at higher redshift in the case of radio-loud quasars (see Table~\ref{Table:mean_fluxes}). A possible suppression of the star-formation due to the radio-jet activity would be in agreement with a model of short-lived episodes of radio-loud states in the life of all AGN. These events are associated with the active nucleus and AGN feedback.

The physical mechanisms responsible for triggering the active AGN phase are still debated. Indeed, it is still poorly understood whether the AGN activity impacts star formation or vice versa. Negative AGN feedback, where the AGN emission is believed to be responsible for gas heating, is necessary in order to explain the strong suppression of star formation especially in the most massive galaxies \citep[e.g.][]{Croton2006,Hopkins2010}. The feedback process becomes more complicated in the case of powerful radio sources where there are results that suggest a positive feedback due to the jets inducing star formation in the host galaxy \citep[e.g.][]{Elbaz2009}. These two mechanisms could be the possible explanation for the star-formation difference between the two populations and the minimum observed in the case of radio-loud quasars. 

We found that the $\left\langle {\rm SFR}\right\rangle $ as a function of optical luminosity shows a bi-modality for $\log_{10}(L_{\rm opt}/{\rm W})\leq38.5$ with the radio-loud quasars covering the upper level. If this bi-modality could be explained by the presence or absence of powerful radio jets, what could explain the same level of star formation for both populations at $\log_{10}(L_{\rm opt}/{\rm W})>38.5$?  As we move to higher optical luminosities, the AGN luminosity increases as a result the direct effect of the radiation from the AGN on the host galaxy ISM. In this case, the feedback is predominantly negative, though occasional positive feedback may occur in the form of jet-induced star formation. As the jets cannot now play the critical role they did at lower luminosities both of the populations have the same star-formation trend. These results are in agreement with our previous work in radio-loud and radio-quiet quasars \citep{Kalfountzou2012}.

\subsection{Host galaxy and dust properties}

Based on diverse studies of several samples, it can be said that radio-loud quasars are associated with luminous elliptical galaxies while radio-quiet quasars are usually found in both elliptical and spiral hosts, depending on the optical luminosity threshold. Generally, it has been proposed that the nuclear luminosity is related to the morphology of the host, but AGN more luminous than a certain luminosity limit can only be hosted by massive spheroidals \citep[e.g.][]{McLure1999,Dunlop2003}. Based on this assumption, our results for different dust temperature could have their origin in the different hosts of radio-loud and radio-quiet quasars.

In the case of the single-temperature model, we found that radio-loud quasars tend to have lower dust temperatures, at least for lower redshifts and/or lower optical luminosities. Low temperatures are associated with the old stellar population of elliptical galaxies. This fact is in agreement with the previously mentioned studies regarding the hosts of radio-loud and radio-quiet quasars. On the other hand, the low dust temperature could be associated with a strong cold component described by the two-temperature grey-body model. Dust temperatures of 10-15 K would imply dust masses of up to $10^{10}{\rm M_{\odot}}$, quite unrealistic for the case of elliptical hosts and generally for quasars' hosts where the expected range of dust mass is $10^{7}-10^{9}{\rm M_{\odot}}$. In our sample, despite the low temperatures just a few sources are found to have $M{\rm _{dust}}>10^{9}{\rm M_{\odot}}$, which is not unexpected as most of them have FIR fluxes even lower than the $2\sigma$ detection limit. Moreover, based on the single-temperature model, we found that both the populations tend to have statistically indistinguishable dust masses. 

An additional point which could play a significant role in the observed differences would be the gas supply in the host galaxies of radio-loud and radio-quiet quasars. The gas content is the fundamental ingredient driving star formation in galaxies. Additionally, AGNs are preferentially hosted by gas rich galaxies \citep[e.g.][]{Silverman2009,Vito2014} which is not surprising since gas accretion onto SMBH is the process at the origin of nuclear activity. Given the dependency of both SFR and AGN on the gas content, the enhanced star formation in AGN galaxies appears to be primarily the result of a larger gas content, with respect to the bulk of the galaxy population at similar stellar masses \citep[e.g.][]{Rosario2012,Santini2012}. Many semi-analytic models and direct observations suggest that the gas fractions in galaxies grow at lower stellar masses and, at fixed mass, increase at earlier cosmic epochs. In the local Universe, low mass galaxies are generally gas-rich and actively star-forming, while the highest mass galaxies are almost always gas-poor and have very little ongoing star formation. This is probably why optical AGN with the highest values of $L/L_{Edd}$ tend to occur in galaxies with the smallest bulges and black holes (Heckman et al 2004). Assuming Gaussian quasar Eddington ratio distributions at all epochs, then the optical luminosity which is used as an AGN activity tracer would map into BH mass and thus on galaxy mass. In this case, radio-loud quasars with lower optical luminosities should, on average, be associated with lower mass and gas-rich galaxies (see Figure~\ref{fig:sum_up}, right panel) for which the effects of a jet-driven star-formation rate may be more evident. On the other hand, the fact that no SFR difference is detected between the two populations at higher redshifts or at higher optical luminosities, when gas fractions should grow, could imply that both populations evolve in gas fractions at the same rate.

In order to explain these possible temperature differences we have to take into account that the integrated dust temperature depends also on the dust distribution throughout the galaxy. Previous studies \citep[e.g.][]{Goudfrooij1995,Leeuw2004} investigating the origin of dust in elliptical galaxies proposed the presence of various components. Similarly, we used a two-component model to describe the FIR properties of our sample, a warm dust component ($T_{w}=35$ K) and a cold one ($T_{c}=15$ K). \cite{Goudfrooij1995} proposed the presence of at least two sources of the observed interstellar matter (ISM) in elliptical galaxies, mass-losing giant stars within the galaxy and galaxy interactions. Minor mergers and/or accretion of material from nearby companions could possible explain the presence of the warm and cold components. Such an assumption of an external origin for the ISM in the early-type galaxies leads to a strong link with the environment of quasars. \cite{Falder2010} showed that radio-loud AGN appear to be found in denser environments than their radio-quiet counterparts at $z \sim 1$. These environments represent ideal candidates for galaxy-galaxy interactions. In this case, the cold dust properties in radio-loud quasars could have an external origin.

\section{CONCLUSIONS} \label{sec:conclusions}

In this paper we have studied the far-infrared properties and the star-formation of matched samples of radio-loud and radio-quiet quasars. The main result of our study is that radio-loud quasars have higher star-formation rates than radio-quiet quasars at low optical luminosities. This result is in agreement with our previous work \citep{Kalfountzou2012} where the [\mbox{O\,{\sc ii}}] emission was used as a tracer of the star-formation. 

Additionally, we have found a strong correlation between jet activity and the star-formation, controlling the effect of redshift, in the case of radio-loud quasars and especially at low optical luminosities and redshifts. This correlation supports the idea of the jet-induced star-formation.

The possible differences we found between the two populations regarding the dust mass and dust temperature could explain the differences in star-formation rate, but they also point the way forwards further investigation of the evolution of their host galaxies and their environment and their correlation with AGN activity.

\section*{Acknowledgments}
The authors would like to thank Micha{\l}~J.~Micha{\l}owski for useful comments and the anonymous referee for a helpful and constructive report. The {\it Herschel}-ATLAS is a project with {\it Herschel}, which is an ESA space observatory with science instruments provided by European-led Principal Investigator consortia and with important participation from NASA. The {\it H}-ATLAS website is \url{http://www.h-atlas.org/}. This work used data from the SDSS DR7. Funding for the SDSS and SDSS-II has been provided by the Alfred P. Sloan Foundation, the Participating Institutions, The National Science Foundation, the U.S. Department of Energy, the National Aeronautics and Space Administration, the Japanese Monbukagakusho, the Max Planck Society and the Higher Education Funding Council for England.

\bibliography{biblio.bib}

\appendix

\section{Synchrotron contamination} 

The far-infrared luminosity is used as a measure of the radiation from dust, which may be heated by star-formation and/or the central quasar nucleus. However, since the radio-loud sample includes high radio flux density sources, it is possible that the far-infrared flux densities we measure may be subject to contamination from synchrotron emission not associated with star formation. The spectra of powerful radio-loud AGN are in some cases entirely dominated by synchrotron emission from the jets at all wavelengths. Radio spectra have been compiled for each radio-loud source, with the aim of subtracting the radio contribution to the FIR emission. 

\begin{figure*}

\begin{center}

\subfigure[Examples of the 9 sources where the extrapolation is massively overestimating the synchrotron contamination. The 9 sources with spectral energy distribution similar to these examples have been rejected from our sample.]{
\hspace{-0.8cm}
\includegraphics[scale=0.6]{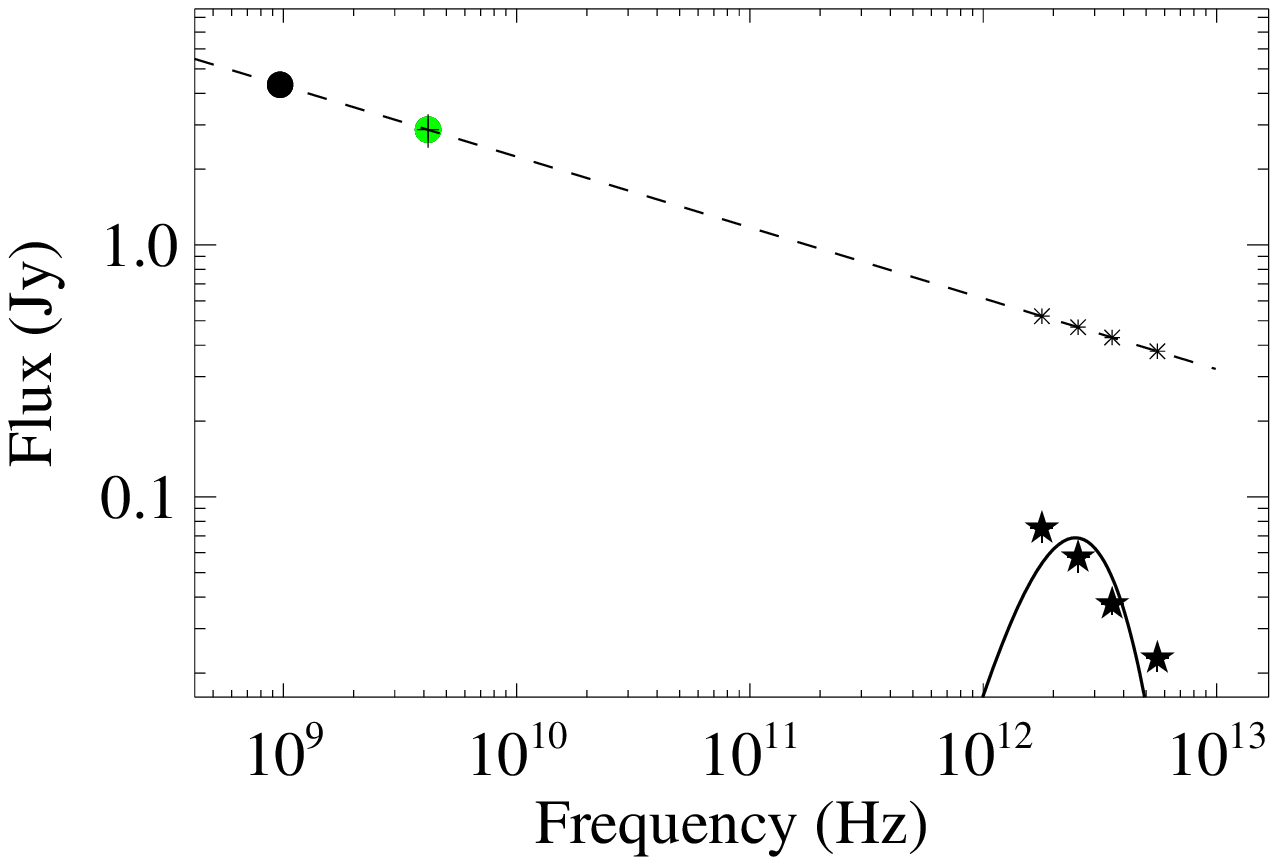}
\hspace{-0.8cm}
\includegraphics[scale=0.6]{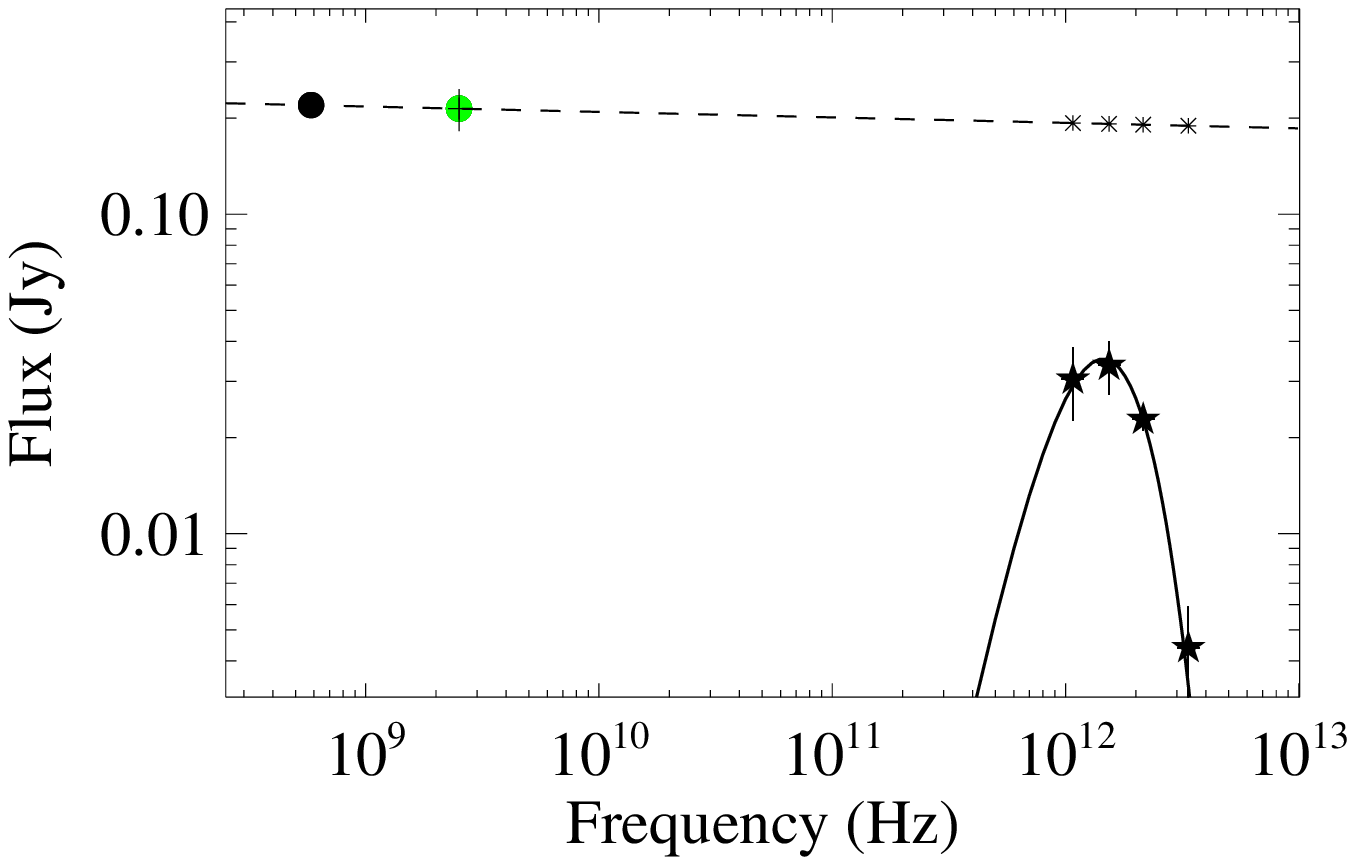}
}

\subfigure[Examples of the 10 sources found having strong synchrotron contamination. The 10 sources with spectral energy distribution similar to these examples have been rejected from our sample.]{
\hspace{-0.8cm}
\includegraphics[scale=0.6]{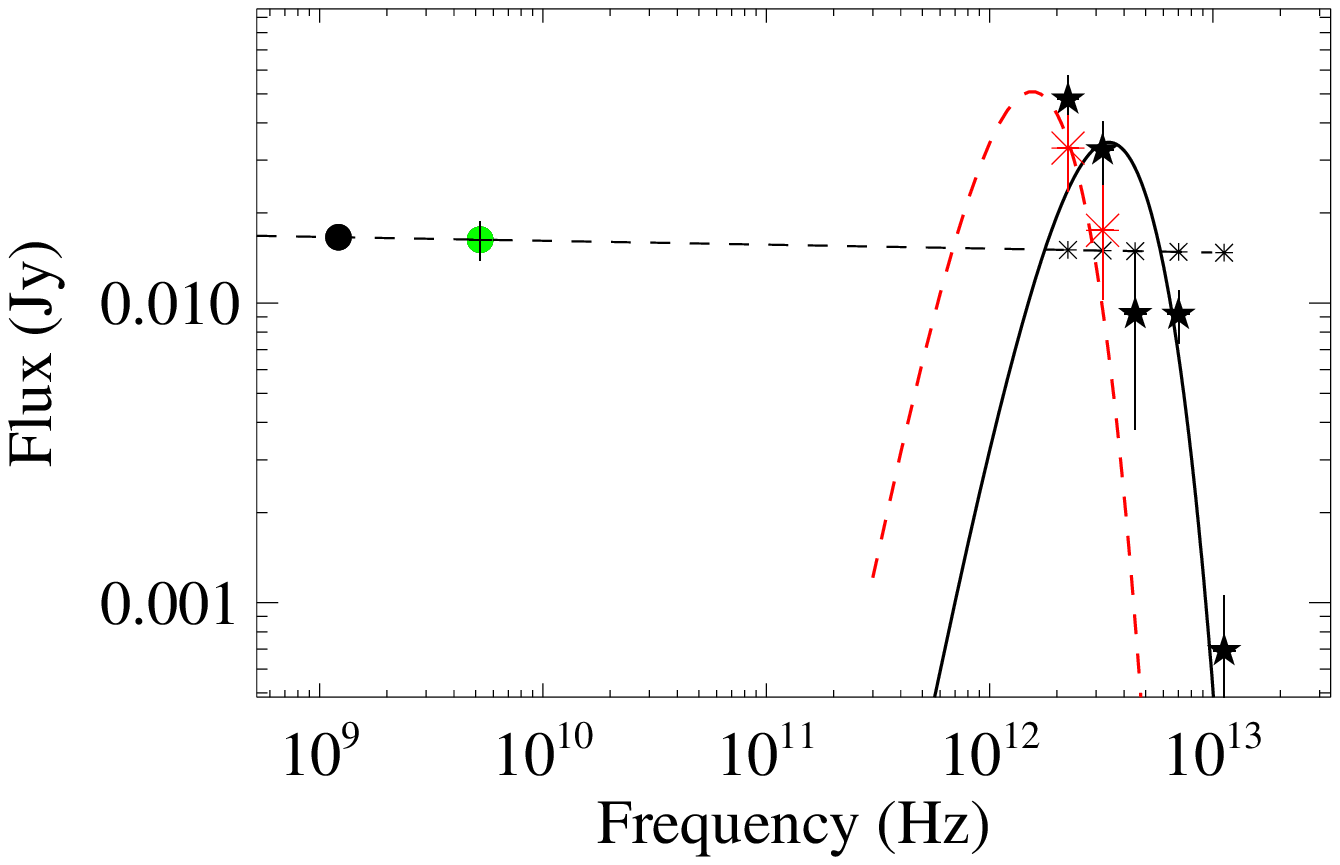}
\hspace{-0.8cm}
\includegraphics[scale=0.6]{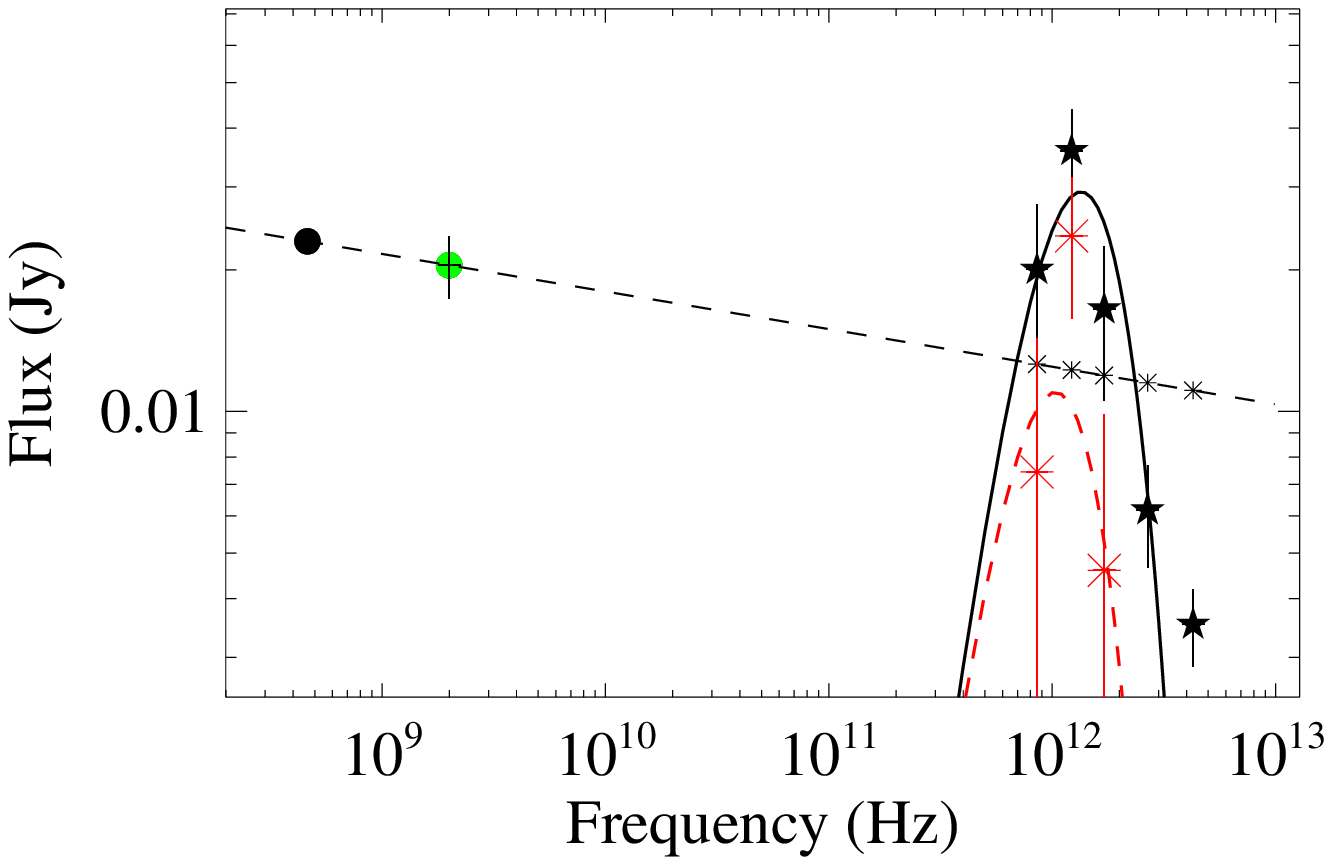}
}

\subfigure[Examples of the 52 sources found having weak or no synchrotron contamination. The sources with spectral energy distribution similar to these examples have been included in our sample. Applying the correction to these sources has no impact on the derived temperatures and luminosities.]{
\hspace{-0.8cm}
\includegraphics[scale=0.6]{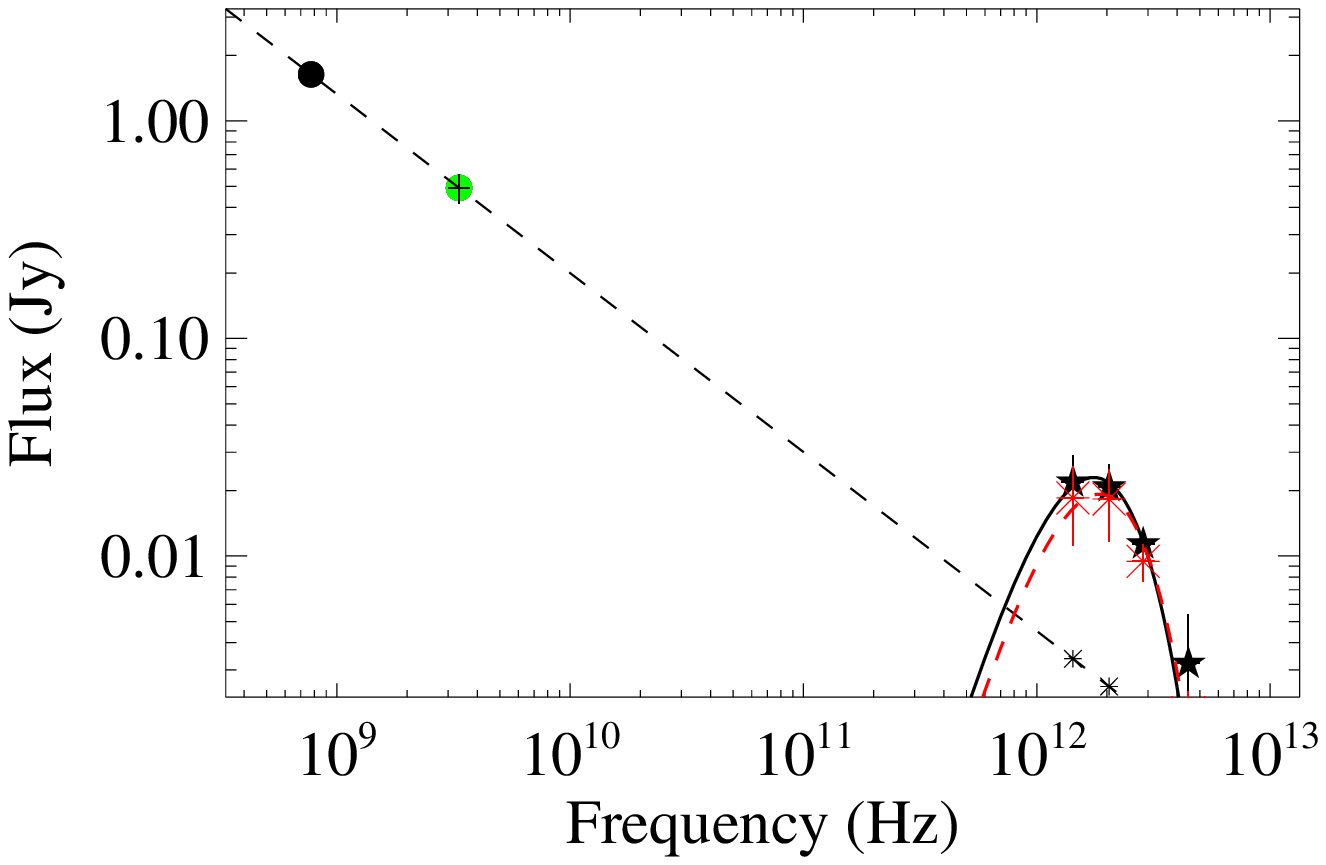}
\hspace{-0.8cm}
\includegraphics[scale=0.6]{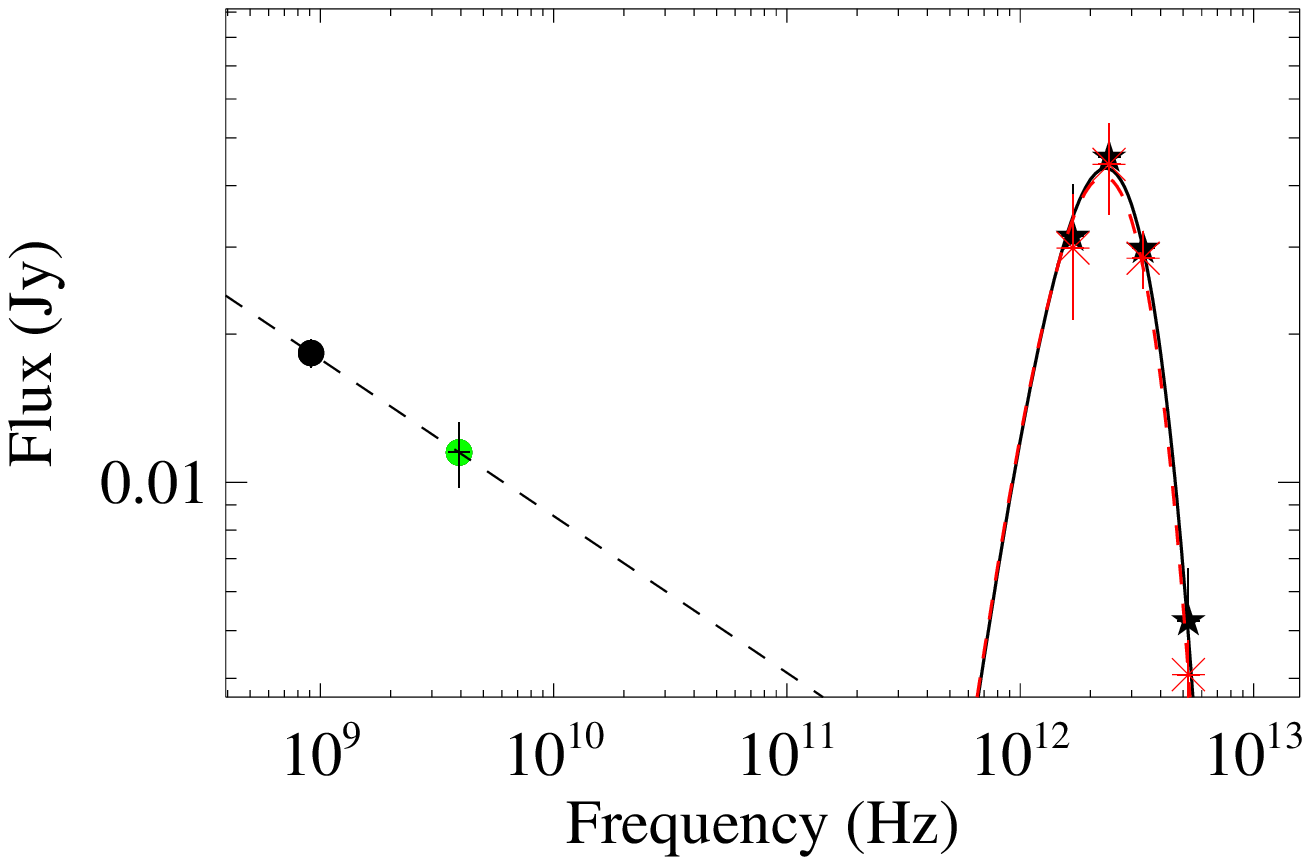}
}

\end{center}

\caption{Spectral energy distribution at radio and FIR wavelengths for a selected sample of radio-loud quasars. \textit{Filled black stars}: the FIR data, \textit{Circles}: the radio data, green for FIRST and black for GMRT, \textit{Black small stars}: synchrotron contamination at SPIRE and PACS bands, \textit{Red asterisks}: the subtracted flux at SPIRE and PACS bands, \textit{Black dashed line}: Linear fit to radio data, \textit{Black solid line}: grey-body fit, \textit{Red dashed line}: grey-body fit after synchrotron subtraction.}
\label{fig:SED_contam}
\end{figure*}

All of the radio-loud quasars in our sample have a detected counterpart in FIRST within a search radius of 5 arcsec. In order to estimate their spectral index we also cross-matched our sample to the Giant Metrewave Radio Telescope (GMRT) catalogue of \cite{Mauch2013}, who have coverage of the H-ATLAS $9^{h}$, $12^{h}$ and $14.5^{h}$ areas at 325 MHz, using a simple positional cross-matching with a maximum of 5 arcsec. Despite the incomplete sky coverage and variable sensitivity of the GMRT survey, a total of 71/141 sources are found to have 325 MHz counterparts. For the matched objects, we can measure their spectral index assuming a power law and then use their mean spectral index for the rest of the population.

In Fig.~\ref{fig:SED_contam} we present a sample of the spectral energy distributions (SEDs) of the radio-loud quasars using the available radio and FIR fluxes. The data include the \textit{Herschel} and the VLA (FIRST catalogue; \citealp{Becker1995}) observations presented in Section \ref{data} and the 325 MHz radio fluxes taken from the GMRT catalogue. Using the extrapolation of the radio fluxes (dashed black line) we attempt to subtract the synchrotron contamination of the FIR fluxes. The subtracted FIR flux densities are fitted with the grey-body model once again (red dashed line) to produce a new estimation of the free parameters.  

In the cases where the subtracted FIR fluxes do not fall close to the original FIR flux densities (within the errors) for more than two FIR bands, the parameters of the new grey-body fitting have changed significantly within the errors from the original ones. In these cases we have found that synchrotron emission strongly affects the FIR flux densities and the FIR luminosity and the sources are rejected from our sample. Specifically, we have divided our sample into 3 categories a) sources where the extrapolation of the radio fluxes massively overestimates the synchrotron contamination (Fig.~\ref{fig:SED_contam}a), b) sources where the synchrotron emission strongly affects the FIR flux densities (Fig.~\ref{fig:SED_contam}b) and sources where the synchrotron contamination is weak and the FIR flux densities are not affected at all (Fig.~\ref{fig:SED_contam}c). 

From our sample of the 71 radio-loud quasars with both FIRST and GMRT radio detections we found 9 sources belong to the (a) category. The examples of two of these sources are presented in Fig.~\ref{fig:SED_contam}a. It is obvious that the straight-line extrapolation of the low-frequency radio emission massively overestimates the synchrotron contamination at the FIR bands;for these sources radio data at higher frequencies would be required in order to describe accurate radio spectra. Due to the lack of high-frequency radio data we had to reject these 10 sources from our sample in order to ensure that the synchrotron emission does not affect the star formation estimation in the radio-loud population. We should mention that only one of these sources has FIR detections at the $3\sigma$ level.

In the second (b) category we have classified the 10 sources with strong synchrotron contamination. Examples of two of these sources are presented in Fig.~\ref{fig:SED_contam}b and they show that all the FIR flux densities appear to be seriously contaminated with non-thermal synchrotron. Although we expect the radio spectra to appear curve at higher frequencies and have less effect on the higher-frequency FIR bands (e.g. PACS bands) it seems that the 500-$\mu$m and 350-$\mu$m detections are likely to be seriously synchrotron contaminated. In order to classify a source as seriously contaminated we compare the results of the grey-body fitting using the original FIR flux densities (black stars) and the FIR flux densities corrected for synchrotron contamination (red stars). As the examples show in Fig.~\ref{fig:SED_contam}b, the grey-body fitting after correction for synchrotron contamination (red dashed line) is significantly different from the original one (black solid line) implying that the parameters estimated using the original grey-body fitting are strongly affected by the synchrotron emission. These 10 sources have been rejected from our sample due to their probably serious contamination from non-thermal synchrotron emission.

In the third (c) category we have classified the remaining 52 sources out of the 71 with both FIRST and GMRT radio detections. The examples of two of these sources are presented in Fig.~\ref{fig:SED_contam}c. In this class are sources with weak (not significant) synchrotron contamination. As the examples show in Fig.~\ref{fig:SED_contam}c, the FIR flux densities after correction for synchrotron contamination (red stars) are within the 1$\sigma$ errors of the original FIR flux densities (black stars) and as a result the estimated parameters from the grey-body fittings (black solid line and red dashed line) using the corrected and the original FIR flux densities are within their errors. All 52 sources with similar SEDs to the examples in Fig.~\ref{fig:SED_contam}c are retained in our sample. 

Overall, we have found 21 objects of our detected at 325 MHz sample where the synchrotron contamination strongly affects the estimates of the grey-body fitting, indicating that these objects have the potential for contamination by their synchrotron components. These sources are rejected from further study. For the rest of the sources which are detected at 325 MHz, we are able to subtract the synchrotron contamination and fit a new grey-body model using the subtracted fluxes. 

Among the rest 70/141 sources that are undetected in the GMRT data, there are 8 sources with available radio data in the literarure \citep{Griffith1995,Douglas1996,Cohen2007,Healey2007,Mason2009} which are used in order to estimate the spectral indiced. One of them shows significant synchrotron contamination and has been removed from the sample. For the other 62 undetected in the GMRT data we conservatively used $\alpha=-0.4$, the minimum value observed in the GMRT-detected sources, to estimate the maximum possible synchrotron contamination. The main characteristic of the sample undetected by GMRT is the faint radio emission at 1.4 GHz, compared to the rest of the radio-loud quasars. The main bulk of these sources has $S_{1.4\rm GHz}<10~{\rm mJy}$ while, the mean value of this sample is $\langle S_{1.4\rm GHz}\rangle= 5.79\pm 0.81~{\rm mJy} $. Due to their faint radio emission, we do not expect for most of them strong contamination. We found that 26 sources show possible synchrotron contamination and they have been removed them from our sample. Finally, our sample consists of 93 radio-loud quasars.

In order to investigate whether there are any particular trends for the sources detected by \textit{Herschel}, we investigated the level of synchrotron contribution in those sources. Due to the limited number of detected radio-loud quasars, we used as a detection limit the 3$\sigma$ level at 250 $\mu$m. We found 26 objects with an available GMRT detection out of the 46 radio-loud quasars with a 3$\sigma$ detection at 250 $\mu$m and as a result, estimated spectral index. In this case, we find a consistent spectral index; $\alpha=0.66\pm0.08$. 

A final method of investigating the synchrotron contamination level is to study the level of core emission. A reasonable estimation of the level of compact emission can be derived from the comparison of the NVSS and FIRST fluxes, investigating whether the quasar radio fluxes are underestimated due to the FIRST survey resolving out extended flux. The cross-match with the NVSS catalogue gave us a total of 90 matches within a 5 arcsec radius. Among these there are 58 sources with a GMRT detection. Comparing the NVSS - FIRST fluxes we found a fraction of $7.0\pm1.7$ per cent excess in their NVSS fluxes. No significant differences were found even when we compared the NVSS - FIRST emission for the sub-samples that are detected and undetected with GMRT. Such a small fraction shows there is no evidence that either the FIRST fluxes or the estimated spectral indices of the sources are underestimated. On the other hand, the low level of extended emission shows that the radio sources are fairly compact and a flatter radio spectrum would be expected. However, a comparison of the spectral index with the NVSS shows no particular trend. 

Overall, we have found that out of the 141 objects in our radio-loud quasar sample, 21 radio-loud quasars have significant non-thermal contamination in their FIR emission while an additional sample of 27 sources possible has strong contamination using an upper limit for their radio spectral index. These objects have been rejected from our sample.
We emphasize that this is a conservative estimate, given that the steep-spectrum synchrotron component is likely to fall more quick than the fitted power-law at higher frequencies due to spectral aging of the electron population. Therefore, our fitting extrapolation is likely to provide an overestimate of the synchrotron contamination at FIR wavelengths in our sample, especially in the cases of power-law fitting. 
Radio data at higher frequencies would give us a clearer view of the possibility of a flat, core dominated spectrum in this frequency range, although our analysis does not support the presence of a flat spectrum at shorter wavelengths.

\section{Two-Temperature model} \label{2D_fitting}

The estimation of the dust mass has been made based on the measured temperature of the grey-body model. Comparing our results with those of \cite{Dunne2011} for {\it Herschel}-detected $z < 0.5$ galaxies, we see that the isothermal dust temperatures we measure span the same range. Taking into account the fact that we use a $\beta=2.0$ emissivity index, our dust mass measurements should increase by $30-50$ per cent from those of \cite{Dunne2011} with the same temperature. Indeed, for a mean temperature of $20-25$ K our population is found to have $\sim 10^{8.0}~{\rm M_{\odot}}$ . One question is if the estimated isothermal dust mass can be biased low, as the dust exists at a range of temperatures in galaxies, while the mass we have estimated is that of the dust close to the source of heating (star-forming regions) which warm it enough to emit at FIR wavelengths. Another important question is whether the presence of a cold component could explain the differences we found for the two populations regarding their dust temperatures and dust masses.

To investigate this we use a model which requires two components of dust. The two required components consist of cold dust with $T_{c}\sim10-25$ K and warmer dust with $T_{w}\sim25-60$ K. The cold component is associated with the old stellar population and the warm one with the current star formation. The luminosity of the warm component is primarily the indicator of the star-formation rate. Previous studies preferred to use two fixed temperatures (a cold and a warm one) in order to fit the two-temperature model. However, the correct choice of the fixed temperatures would be difficult as our single-temperature results show that the two populations (radio-loud and radio-quiet quasars) may have different dust temperatures. In order to overcome this problem of the possible different temperatures between the two populations, we fit a two-temperature model for several different temperature pairs within $\pm5$ K of our initial chosen fixed temperatures, $T_{c}=15$ K and $T_{w}=35$ K. 

Using each possible pair of cold and warm component temperatures we estimate the FIR luminosities and the dust mass for each component. For the two-component model the FIR luminosity is:
\begin{equation}
\centering
{\rm LFIR}=N_{w}\nu^{\beta}B_{\nu}\left(\frac{\nu}{1+z},T_{w}\right)+N_{c}\nu^{\beta}B_{\nu}\left(\frac{\nu}{1+z},T_{c}\right)
\end{equation}
where $N_{w}$ and $N_{c}$ are the relative contribution due to the warm and cold dust components. The dust mass is computed from the sum of the masses in the two temperature components \citep{Vlahakis2005}: 
\begin{equation}
\centering
M_{\rm d}=\frac{L_{250}}{4\pi \kappa_{250}}
\left[\frac{N_{w}}{B_{\nu}\left(\frac{\nu}{1+z},T_{w}\right)}+\frac{N_{c}}{B_{\nu}\left(\frac{\nu}{1+z},T_{c}\right)}\right]\frac{1}{N_{w}+N_{c}}
\end{equation}
where $\kappa_{250}=0.89~{\rm m^{2}~kg^{-1}}$ is the dust mass absorption coefficient and $B_{\nu}$ is the two-temperature modified Planck function.

In the cases where the objects are well described by a single temperature for the warm component that is significantly different from $T_{w}=35~{\rm K}$ the two-temperature model with fixed temperatures fit less well.
However, we have found a good correlation between the FIR luminosities of the two fitting models. In contrast, the estimated dust masses show less good agreement with higher scatter. This suggests that, at least for this sample, the estimation of the FIR luminosity is not strongly affected by the fitting model, while the dust mass must be interpreted with a little more care. Comparing the contamination of the cold component to the total FIR and $250~\mu$m luminosities we found that in both populations the warm component dominated the overall luminosity at a higher level than 70 per cent. This result shows that any differences found should not be a result of a strong cold component in any of the two populations. 

\bsp

\label{lastpage}

\end{document}